\documentclass{ws-ijmpd}
\usepackage[super,compress]{cite}
\usepackage{graphicx}
\usepackage{amsmath}
\usepackage{amsfonts}
\usepackage[percent]{overpic}
\usepackage{enumitem}
\usepackage{float}
\begin{document}

\markboth{Hadyan L. Prihadi, Muhammad F. A. R. Sakti, Getbogi Hikmawan, Freddy P. Zen}
{Dynamics of Charged and Rotating NUT Black Holes in Rastall Gravity}

%
\catchline{}{}{}{}{}
%

\title{Dynamics of Charged and Rotating NUT Black Holes in Rastall Gravity}

\author{Hadyan L. Prihadi$^{1,a}$, Muhammad F. A. R. Sakti$^{1,b}$, Getbogi Hikmawan$^{1,2c}$, and Freddy P. Zen$^{1,2d}$}

\address{$^1$Theoretical High Energy Physics, Department of Physics\\
Institut Teknologi Bandung, Jl. Ganeca 10 Bandung, Indonesia.}
\address{$^2$ Indonesia Center for Theoretical and Mathematical Physics (ICTMP)\\
Institut Teknologi Bandung, Jl. Ganesha 10 Bandung,
	40132, Indonesia.}
\address{$^a$hadyanluthfanp@students.itb.ac.id, $^b$m.fitrah@students.itb.ac.id\\ $^c$getbogihikmawan@fi.itb.ac.id, $^d$fpzen@fi.itb.ac.id}

\maketitle

\begin{history}
\received{Day Month Year}
\revised{Day Month Year}
\end{history}

\begin{abstract}
In this work, we generalized the Kerr-Newman-NUT black hole solution in Rastall gravity from Ref. \citen{1}. Here we are more focused on the black hole dynamics such as the event horizons, ergosurface, ZAMO, thermodynamic properties, and the equatorial circular orbit around the black hole such as static radius limit, null equatorial circular orbit, and innermost stable circular orbit. We present how the NUT and Rastall parameter affects the dynamic of the black hole.
\end{abstract}


\ccode{PACS numbers:}


\section{Introduction}
In general relativity, Einstein field equation (EFE) gives many solutions that represent a geometrical structure of space-time. Many EFE solutions give rise to an object called the black hole, \textit{i.e.} a region of space-time that possesses a coordinate singularity called event horizon. Some of the black hole solutions that have been found are the static and spherically symmetric Schwarzschild black hole, the rotating Kerr black hole, the charged Reissner-Nordstr$\ddot{\text{o}}$m black hole, and the rotating and charged Kerr-Newman black hole. Decades after the formulation of general relativity, the solutions to Einstein's equations have been found to be even more numerous, \textit{e.g.} the black hole solution surrounded by Quintessence field, \textit{i.e.} a model of a scalar field that explains the expansion of the universe by governing a negative pressure\cite{6,32}, the Kiselev black hole \cite{7} and its rotating counterparts \cite{8,9}. Other studies involving scalar field in cosmological scale can also be found in Ref. \citen{33}. There also exists a black hole solution form Einstein-Maxwell-Dilaton theory\cite{10} or even its higher dimensional extension\cite{11}, and many more. Aside from the mathematical formulations, a black hole has also been found as a real astronomical object recently, \textit{e.g.} the observation of gravitational waves from a merging of two black holes by LIGO\cite{12}, and the first ever real black hole image captured by the event horizon telescope \cite{18}.\\
\indent Other interesting black hole solutions is the one that came from a modified theory of gravity, such as Rastall gravity\cite{2} which said that theory of general relativity does not need to always  fulfill the conservation of the energy-momentum tensor $\nabla^{\mu}T_{\mu\nu}=0$, but in general it could be $\nabla^{\mu}T_{\mu\nu}=\lambda a_{\nu}$ for some vector $a_{\nu}$ and a proportional constant $\lambda$ and we recover the usual conservation law when $\lambda\rightarrow0$. Since the energy-momentum tensor is related to the space-time curvature by the Einstein field equation, $a_{\nu}$ also needs to be related to the space-time curvature, therefore the most suitable result is $a_{\nu}=\nabla_{\nu}R$, where $R$ is the Ricci scalar. Various black hole solutions in Rastall gravity have been found, \textit{e.g.} the static black hole\cite{24} and the charged rotating black hole solutions\cite{25}. Similar works involved in Rastall background recently can be found in Ref. \citen{26}.\\
\indent To make the solution more general, here we add the NUT parameter \cite{41,42} to a charged rotating black hole solution in Rastall gravity. The NUT parameter can be interpreted as the \textit{twisting} parameter of the space-time \cite{43} or the \textit{gravito-magnetic} monopole \cite{44}. However, the physical interpretation of the NUT parameter is still debatable until now. Similar work has been done in Ref. \citen{1}, but here we give even more general result, by considering the $\theta$ dependence of the horizon, and focused more in the thermodynamic stability and the geodesic of a particle around the black hole.\\
\indent In this work, we begin with obtaining the static and spherically symmetric solution in Rastall gravity, and generate the rotation and NUT parameter using the Newman-Janis algorithm \cite{45} in section 2, and we investigate the horizon and ergoregion structure in section 3, as well as the zero angular momentum observer (ZAMO). Thermodynamic properties are investigated in section 4, and its thermodynamic stability are also studied. In section 5, we worked on the equatorial circular orbit of a time-like and null-like particle, \textit{i.e.} to study its static radius limit, null equatorial circular orbit, and the innermost stable circular orbit (ISCO). Finally, in section 6, we summed up all of the topics with concluding remarks.

\section{Kerr-Newman-NUT Solution in Rastall Gravity}

In this section, we introduce the metric solution for a rotating and charged black hole with NUT parameter in Rastall gravity. Based on Rastall hypothesis which stated that the energy-momentum tensor $T_{\mu\nu}$ in general does not always satisfy the conservation law condition $\nabla^{\mu}T_{\mu\nu}=0$, but rather it should satisfy \cite{2}
\begin{equation}
\nabla^{\mu}T_{\mu\nu}=\lambda\nabla_{\nu}R\label{eq:1}
\end{equation}
where $\lambda$ is the proportional constant that measures the deviation from Einstein theory of gravity. We can clearly see that if $\lambda\rightarrow0$, it recovers the conservation law of the energy-momentum tensor. Again, when we work in flat space-time, the usual conservation law is recovered. With that condition, the field equation becomes
\begin{equation}
G_{\mu\nu}+\kappa\lambda Rg_{\mu\nu}=\kappa T_{\mu\nu},\label{eq:2}
\end{equation}
where $G_{\mu\nu}=R_{\mu\nu}-\frac{1}{2}Rg_{\mu\nu}$ is the usual Einstein tensor, $\lambda$ is then called the Rastall parameter, and $\kappa=\frac{8\pi G}{c^4}$ is the proportional constant in Einstein field equation that connects Einstein Tensor with energy-momentum tensor. This equation of motion will be named the Einstein-Rastall field equation hereafter.
\subsection{Static and Spherically Symmetric Solution in Rastall Gravity}
\indent To find the solution for a spherically symmetric space-time in Rastall gravity, we consider the Schwarzschild-like solution of the metric as
\begin{equation}
ds^2=-f(r)dt^2+\frac{dr^2}{f(r)}+r^2d\Omega^2, \label{eq:3}
\end{equation}
where $f(r)$ is a function that depends only on coordinate $r$ that will be obtained by solving the Einstein-Rastall field equation in Eq. (\ref{eq:2}) with some properly chosen energy-momentum tensor. Here $d\Omega^2=d\theta^2+\sin^2\theta d\phi^2$ implies the two dimensional sphere part of the metric. By plugging the metric from Eq. (\ref{eq:3}) to the Einstein-Rastall field equation in Eq. (\ref{eq:2}), we have
\begin{align}
H^t_t&\equiv G^t_t + \kappa\lambda R=\frac{1}{r^2}(f'r-1+f)+\kappa\lambda R, \label{eq:4}\\
H^r_r&\equiv G^r_r + \kappa\lambda R=\frac{1}{r^2}(f'r-1+f)+\kappa\lambda R,\\
H^{\theta}_{\theta}&\equiv G^{\theta}_{\theta} + \kappa\lambda R=\frac{1}{r^2}\bigg(rf'+\frac{1}{2}r^2f''\bigg)+\kappa \lambda R,\\
H^{\phi}_{\phi}&\equiv G^{\phi}_{\phi} + \kappa\lambda R=\frac{1}{r^2}\bigg(rf'+\frac{1}{2}r^2f''\bigg)+\kappa\lambda R, \label{eq:7}
\end{align}
where $R$ is the Ricci scalar for the metric in Eq. (\ref{eq:3}), given by
\begin{equation}
R=-\frac{1}{r^2}(r^2f''+4rf'-2+2f).\nonumber
\end{equation}
Here we use the prime notation as the derivative with respect to coordinate $r$, $f'\equiv df/dr$. From Eq. (\ref{eq:4}) to Eq. (\ref{eq:7}), we can see that $H^t_t=H^r_r$ and $H^{\theta}_{\theta}=H^{\phi}_{\phi}$. It indicates that we have to look for the energy-momentum tensor that satisfies both $T^t_t=T^r_r$ and $T^{\theta}_{\theta}=T^{\phi}_{\phi}$. We will see that the electromagnetic and quintessence energy-momentum tensor is a perfect match and the total energy-momentum tensor will read as
\begin{equation}
T_{\mu\nu}=\mathcal{E}_{\mu\nu}+\mathcal{J}_{\mu\nu},
\end{equation}
where $\mathcal{E}_{\mu\nu}$ is the electromagnetic energy-momentum tensor, and $\mathcal{J}_{\mu\nu}$ is for the quintessence part.\\
\indent The electromagnetic part of the energy-momentum tensor is given by \cite{1,25}
\begin{equation}
\mathcal{E}_{\mu\nu}=\frac{2}{\kappa}\bigg(F_{\mu\alpha}F^{\alpha}_{\nu}-\frac{1}{4}F_{\alpha\beta}F^{\alpha\beta}g_{\mu\nu}\bigg),
\end{equation}
where $F_{\mu\nu}=\partial_{\mu}A_{\nu}-\partial_{\nu}A_{\mu}$ is the usual electromagnetic field strength tensor with the electromagnetic vector potential $A_{\mu}$, that satisfies Maxwell Equation $\nabla_{\mu}F^{\mu\nu}=0$ and Bianchi's identity $\partial_{[\sigma}F_{\mu\nu]}=0$. Here we only interested in the electrostatic charge $Q$ in the radial direction that gives rise to
\begin{equation}
A_{\mu}dx^{\mu}=-\frac{Q}{r}dt\;\;\;\;\;\rightarrow\;\;\;\;\;F^{tr}=\frac{Q}{r^2}.
\end{equation}
From the equation above, the non-zero part of the electromagnetic energy-momentum tensor is given by
\begin{align}
\mathcal{E}^{t}_t=\mathcal{E}^r_r=-\frac{Q^2}{\kappa r^4}\;\;\;\;\;\text{and}\;\;\;\;\;\mathcal{E}^{\theta}_{\theta}=\mathcal{E}^{\phi}_{\phi}=\frac{Q^2}{\kappa r^4}.\label{eq:11}
\end{align}
We see that Eq. (\ref{eq:11}) satisfies the needed condition for the energy-momentum tensor.\\
\indent Now consider the energy-momentum tensor from \cite{1,25} that also obey the previous Einstein-Rastall symmetry properties as the quintessence part that reads
\begin{equation}
\mathcal{J}^t_t=\mathcal{J}^r_r=-\rho_q(r)\;\;\;\;\;\text{and}\;\;\;\;\;\mathcal{J}^{\theta}_{\theta}=\mathcal{J}^{\phi}_{\phi}=\frac{1}{2}(1+3\omega)\rho_q(r), \label{eq:12}
\end{equation}
where $\rho_q(r)$ is the perfect fluid density and $\omega$ is the parameter from the equation of state that relates pressure with density. Then after plugging the known total energy-momentum tensor from Eqs. (\ref{eq:11}) and (\ref{eq:12}) to the Einstein-Rastall field equation in Eq. (\ref{eq:2}), we will have two differential equations:
\begin{equation}
\frac{1}{r^2}(f'r-1+f)-\frac{\kappa\lambda}{r^2}(r^2f''+4rf'-2+2f)=-\kappa\rho-\frac{Q^2}{r^4}\label{eq:13}
\end{equation}
and
\begin{equation}
\frac{1}{r^2}\bigg(rf'+\frac{1}{2}r^2f''\bigg)-\frac{\kappa\lambda}{r^2}(r^2f''+4rf'-2+2f)=\frac{\kappa}{2}\rho(3\omega+1)+\frac{Q^2}{r^4}.\label{eq:14}
\end{equation}
To solve those equations, we need to consider that we want to recover Reissner-Nordstr$\ddot{\text{o}}$m solution, the spherically symmetric black hole with electric (and magnetic) charge, when both quintessence energy-momentum and $\lambda$ vanishes since that condition gives rise to the Einstein-Maxwell field equation. Therefore, we can use the function
\begin{equation}
f(r)=1-\frac{2M}{r}+\frac{Q^2}{r^2}+h(r) \label{eq:15}
\end{equation}
as the proper ansatz.\\
\indent Our next job is to find the unknown function $h(r)$, which will be much easier to compute. We can also take the function $h(r)$ in the form of $h(r)=-\alpha/r^{\zeta}$ for some constant $\alpha$ and $\zeta$ will be a function of $\lambda$ and $\omega$. After solving Eqs. (\ref{eq:13}) and (\ref{eq:14}) with the ansatz given by Eq. (\ref{eq:15}), we will get the metric for a spherically symmetric charged black hole with quintessence in Rastall gravity as
\begin{equation}
ds^2=-\bigg(1-\frac{2M}{r}+\frac{Q^2}{r^2}-\frac{N_s}{r^{\zeta}}\bigg)dt^2+\frac{dr^2}{\bigg(1-\frac{2M}{r}+\frac{Q^2}{r^2}-\frac{N_s}{r^{\zeta}}\bigg)}+r^2d\Omega^2,\label{eq:16}
\end{equation}
where
\begin{equation}
\zeta=\frac{1+3\omega-6\kappa\lambda(1+\omega)}{1-3\kappa\lambda(1+\omega)}. \label{eq:16'}
\end{equation}
Here we take the constant $\alpha=N_s$ and interpret them as the quintessential density \cite{1}. We can also see that metric solution (\ref{eq:16}) reduces to Kiselev and Reissner-Nordstr$\ddot{\text{o}}$m black hole when $\lambda\rightarrow 0$ and $N_s\rightarrow 0$, respectively.\\
\indent Besides the radial function $f(r)$, we can also solve for the perfect fluid density $\rho_q(r)$ from Eqs. (\ref{eq:13}) and (\ref{eq:14}) as in \cite{1,25}
\begin{equation}
\rho_q(r)=-\frac{3\mathcal{W}_sN_s}{\kappa r^{(2\zeta-3\omega+1)}},
\end{equation}
where
\begin{equation}
\mathcal{W}_s=-\frac{(1-4\kappa\lambda)(\kappa\lambda(1+\omega)-\omega)}{(1-3\kappa\lambda(1+\omega))^2}.
\end{equation}
By definition, the perfect fluid density $\rho_q(r)$ needs to be non-negative. Thus, we restrict $\mathcal{W}_sN_s$ to be a negative value, \textit{i.e.} $\mathcal{W}_sN_s\leq0$. In this paper, we consider some values of $\omega$ such as dust field ($\omega=0$), radiation field ($\omega=1/3$), quintessence field ($\omega=-2/3,-1/3$), and cosmological constant ($\omega=-1$). For surrounding dust and quintessence ($\omega=-2/3$) field, we restrict the Rastall parameter to $0\leq\kappa\lambda\leq1/4$ for a positive value of $N_s$.
\indent  
\subsection{Generating Rotation and NUT Parameter Using Newman-Janis Algorithm}
After obtaining the metric solution for a static, spherically symmetric black hole in Rastall gravity, we now generate the rotating and NUT parameter using the previously well-known process called Newman-Janis algorithm \cite{45}. Briefly, the Newman-Janis algorithm contains 4 steps: Eddington-Finkelstein coordinate transformation, complex transformation, complexification, and Boyer-Lindquist coordinate transformation. \\
\indent First, we perform the Eddington-Finkelstein coordinate transformation $dt=du+f(r)^{-1}dr$ from the metric solution (\ref{eq:16}), such that the metric can be written as the form of the advanced null coordinate
\begin{equation}
ds^2=-f(r)du^2-2dudr+r^2d\Omega^2.
\end{equation}
Next, we write the inverse metric in terms of complex null tetrad $\{e_{(a)}^{\mu}\}=(l^{\mu},n^{\mu},m^{\mu},\bar{m}^{\mu})$ that form the metric tensor component $g^{\mu\nu}=l^{\mu}n^{\nu}+l^{\nu}n^{\mu}-m^{\mu}\bar{m}^{\nu}-m^{\nu}\bar{m}^{\mu}$. The complex null tetrad are
\begin{align}
l^{\mu}&=\delta^{\mu}_r,\label{eq:20}\\
n^{\mu}&=\delta^{\mu}_u-\frac{1}{2}f\delta^{\mu}_r,\label{eq:21}\\
m^{\mu}&=\frac{1}{\sqrt{2}r}\bigg(\delta^{\mu}_{\theta}+\frac{i}{\sin\theta}\delta^{\mu}_{\phi}\bigg),\label{eq:22}\\
\bar{m}^{\mu}&=\frac{1}{\sqrt{2}r}\bigg(\delta^{\mu}_{\theta}-\frac{i}{\sin\theta}\delta^{\mu}_{\phi}\bigg).\label{eq:23}
\end{align}
Equation (\ref{eq:20}) to (\ref{eq:21}) obey the orthonormal condition for complex null tetrads, \textit{i.e.} all the contractions are zero except for $l^{\mu}n_{\mu}=-1$ and $m^{\mu}\bar{m}_{\mu}=1$. Following Newman-Janis prescription, the complex transformations that generate the rotating and NUT parameters are \cite{45}
\begin{align}
\tilde{u}&=u-ia\cos\theta+i2n\ln\sin\theta,\\
\tilde{r}&=r+ia\cos\theta-in,\\
\tilde{M}&=M+in.
\end{align}
Here $M$ is the black hole mass, $a$ and $n$ are interpreted as the rotating and NUT parameter respectively and it is quite common to consider $a$ as the angular momentum per unit mass of the black hole, \textit{i.e.} $a=J/M$. Following this complex transformations, the general vector transformation is
\begin{equation}
x^{\mu}\rightarrow x'^{\mu}=\delta^{\mu}_u+ia\cos\theta(\delta^{\mu}_r-\delta^{\mu}_u)+in(2\ln\sin\theta\delta^{\mu}_u-\delta^{\mu}_r)+\delta^{\mu}_r+\delta^{\mu}_{\theta}+\delta^{\mu}_{\phi},
\end{equation}
and using the transformation $\{e_{(a)}'^{\mu}\}=\frac{\partial x'^{\mu}}{\partial x^{\nu}}\{e_{(a)}^{\mu}\}$, the complex null tetrads become
\begin{align}
\tilde{l}^{\mu}&=\delta^{\mu}_r,\label{eq:28}\\
\tilde{n}^{\mu}&=\delta^{\mu}_u-\frac{1}{2}\tilde{f}(r,\theta)\delta^{\mu}_r,\label{eq:29}\\
\tilde{m}^{\mu}&=\frac{1}{\sqrt{2}\tilde{r}}\bigg(\delta^{\nu}_{\theta}+ia\sin\theta(\delta^{\mu}_u-\delta^{\mu}_r)+i2n\frac{\cos\theta}{\sin\theta}\delta^{\mu}_u+\frac{i}{\sin\theta}\delta^{\mu}_{\phi}\bigg),\label{eq:30}\\
\tilde{\bar{m}}^{\mu}&=\frac{1}{\sqrt{2}\tilde{r}}\bigg(\delta^{\nu}_{\theta}-ia\sin\theta(\delta^{\mu}_u-\delta^{\mu}_r)-i2n\frac{\cos\theta}{\sin\theta}\delta^{\mu}_u-\frac{i}{\sin\theta}\delta^{\mu}_{\phi}\bigg).\label{eq:31}
\end{align}
We see that the previous radial function $f(r)$ becomes a functions $r$ and $\theta$ after complexification $f(r)\rightarrow\tilde{f}(r,\theta)\in\mathbb{R}$ by following rules
\begin{align}
r&\rightarrow\frac{1}{2}(\tilde{r}+\bar{\tilde{r}})=\text{Re}\tilde{r},\\
\frac{M}{r}&\rightarrow\frac{1}{2}\bigg(\frac{\tilde{M}}{\tilde{r}}+\frac{\bar{\tilde{M}}}{\bar{\tilde{r}}}\bigg)=\frac{\text{Re}(\tilde{M}\tilde{r})}{|\tilde{r}|^2},\\
r^2&\rightarrow|\tilde{r}|^2,
\end{align}
and the function $\tilde{f}(r,\theta)$ becomes
\begin{align}
\tilde{f}(r,\theta)=1-\frac{2Mr+2n(-a\cos\theta+n)-Q^2}{\Sigma(r,\theta)}-\frac{N_s}{\Sigma(r,\theta)^{\frac{\zeta}{2}}}
\end{align}
where $\Sigma(r,\theta)$ is defined as $\Sigma(r,\theta)\equiv|\tilde{r}|^2=r^2+(a\cos\theta-n)^2$. Using complex null tetrad from Eq. (\ref{eq:28}) to Eq. (\ref{eq:31}), the non-zero component for the metric tensor with upper indices are
\begin{align}
\tilde{g}^{uu}&=\frac{a^2\sin^2\theta}{\Sigma(r,\theta)}+4n^2\frac{\cos^2\theta}{\Sigma(r,\theta)\sin^2\theta}+4an\frac{\cos\theta}{\Sigma(r,\theta)},\nonumber\\
\tilde{g}^{u\phi}&=\frac{a}{\Sigma(r,\theta)}+2n\frac{\cos\theta}{\Sigma(r,\theta)\sin^2\theta},\;\;\;\tilde{g}^{\phi\phi}=\frac{1}{\Sigma(r,\theta)\sin^2\theta},\nonumber\\
\tilde{g}^{ur}&=-1-\frac{a^2\sin^2\theta}{\Sigma(r,\theta)}-2an\frac{\cos\theta}{\Sigma(r,\theta)},\nonumber\\
\tilde{g}^{rr}&=\tilde{f}+\frac{a^2\sin^2\theta}{\Sigma(r,\theta)},\;\;\;\tilde{g}^{r\phi}=-\frac{a}{\Sigma(r,\theta)},\;\;\;\tilde{g}^{\theta\theta}=\frac{1}{\Sigma(r,\theta)}.\nonumber
\end{align}
Straightforwardly from the metric tensor component after complexification, the new line element in advanced null coordinate is
\begin{align}
ds^2=&-\tilde{f}(r,\theta)du^2-(2a(1-\tilde{f}(r,\theta))\sin^2\theta-4\tilde{f}(r,\theta)n\cos\theta)dud\phi\\\nonumber
&+2(a\sin^2\theta+2n\cos\theta)drd\phi+\Sigma(r,\theta) d\theta^2-2dudr+\sin^2\theta\\\nonumber
&\times\bigg[\Sigma(r,\theta)+(1-\tilde{f}(r,\theta))\times 4an\cos\theta\nonumber+a^2(2-\tilde{f}(r,\theta))\sin^2\theta\\\nonumber
&-\tilde{f}(r,\theta)4n^2\frac{\cos^2\theta}{\sin^2\theta}\bigg]d\phi^2.
\end{align}
Now we are ready to perform the final part of the whole Newman-Janis algorithm: Boyer-Lindquist coordinate transformation. We choose the following transformations to eliminate all the crossing terms and leave $dtd\phi$ term to preserve the axial symmetry. The coordinate transformations are
\begin{align}
du=dt+\psi dr,\;\;\;d\phi=d\phi+\Xi dr,
\end{align}
where $\psi$ and $\Xi$ are undetermined functions. After making all the off-diagonal metric elements vanish except $g_{t\phi}$, we get the expression for $\psi$ and $\Xi$ and it turns out that they are both functions of $r$ and $\theta$ as
\begin{align}
\psi(r,\theta)&=\frac{-(\Sigma(r,\theta)+a^2\sin^2\theta+2an\cos\theta)}{(\tilde{f}(r,\theta)\Sigma(r,\theta)+a^2\sin^2\theta)},\\
\Xi(r,\theta)&=\frac{-a}{(\tilde{f}(r,\theta)\Sigma(r,\theta)+a^2\sin^2\theta)}.
\end{align}
The $\theta$ dependence might rises because we are working with non-vacuum surrounding and modified theory of gravity\cite{8,25}. From this transformations, we finally to the final result of this section: the Kerr-Newman-NUT black hole in Rastall gravity metric solution in Boyer-Lindquist coordinate,
\begin{align}
ds^2&=-\frac{\Delta-a^2\sin^2\theta}{\Sigma}dt^2+\frac{2}{\Sigma}(\Delta\chi-a(\Sigma+a\chi)\sin^2\theta)dtd\phi\nonumber\\
&\;\;\;\;\;+\frac{1}{\Sigma}((\Sigma+a\chi)^2\sin^2\theta-\chi^2\Delta)d\phi^2+\frac{\Sigma}{\Delta}dr^2+\Sigma d\theta^2,\label{eq:40}
\end{align}
where $\Delta=r^2+a^2-n^2-2Mr+Q^2-N_s\Sigma^{\frac{2-\zeta}{2}}$ and $\chi=a\sin^2\theta+2n\cos\theta$ are both functions of $r$ and $\theta$. This solution is similar to the ordinary Kerr-Newman-NUT black hole \cite{46,47,48,49,50,51}, but with extra terms in the $\Delta$ function that arises from quintessence and Rastall parameter. Furthermore, metric solution in Eq. (\ref{eq:40}) will be named the KNN-R solution for short.\\
\indent KNN-R solution also matches previous known black hole solutions, \textit{e.g.} if the Rastall parameter vanishes ($\lambda\rightarrow0$), the solution will become Kerr-Newman-NUT black hole surrounded by quintessential matter, and when both $\lambda,N_s\rightarrow0$, we recover the Kerr-Newman-NUT solution. The Kerr-Newman and Kerr solution arises when $N_s,n\rightarrow0$ and $N_s,n,Q\rightarrow0$, respectively. Finally, we arrived to the static and spherically symmetric Schwarzschild solution when $N_s,n,Q,a\rightarrow0$.\\
\indent Black hole solution in (\ref{eq:40}) is quite interesting because we have the horizon as a function of $\theta$. It indicates that the horizon structure will not be a perfect sphere, as we will discuss more detail in the next chapter. At first, the dependence of $\theta$ seems quite unphysical. But we can take it because that arises from the external properties of the black hole such as the surrounding matter and Rastall parameter, \textit{e.g.} when the Quintessence part vanishes ($N_s\rightarrow 0$), we wipe out any $\theta$ dependence of the horizon and return to the Kerr-Newman-NUT solution with a perfectly spherical horizon.

\section{Horizon and Ergoregion structure}
In this section, we study the horizon and ergoregion structure of the KNN-R black hole that depends on the values of $M,a,Q,n,N_s,\omega,$ and $\lambda$. Finding a solution for the horizon and ergoregion is by determining the root(s) of $\Delta$ and $g_{tt}$, respectively. Occasionally, we are able to solve it analytically but it often needs numerical computation since we are dealing with a nonlinear equation. Furthermore, we will also see that it is possible to have more than one horizons. Because $g^{rr}=\Delta=0$ at the horizon, we will have $g^{\mu\nu}\sigma^{(r)}_{\mu}\sigma^{(r)}_{\nu}=0$, where $\sigma^{(r)}_{\mu}$ is a normal of the horizon's hypersurface and it becomes a null vector. Hence, the horizon is a null hypersurface. 
\subsection{Black Hole Horizon}
\begin{figure}
	\centering
	\begin{overpic}[scale=0.3]{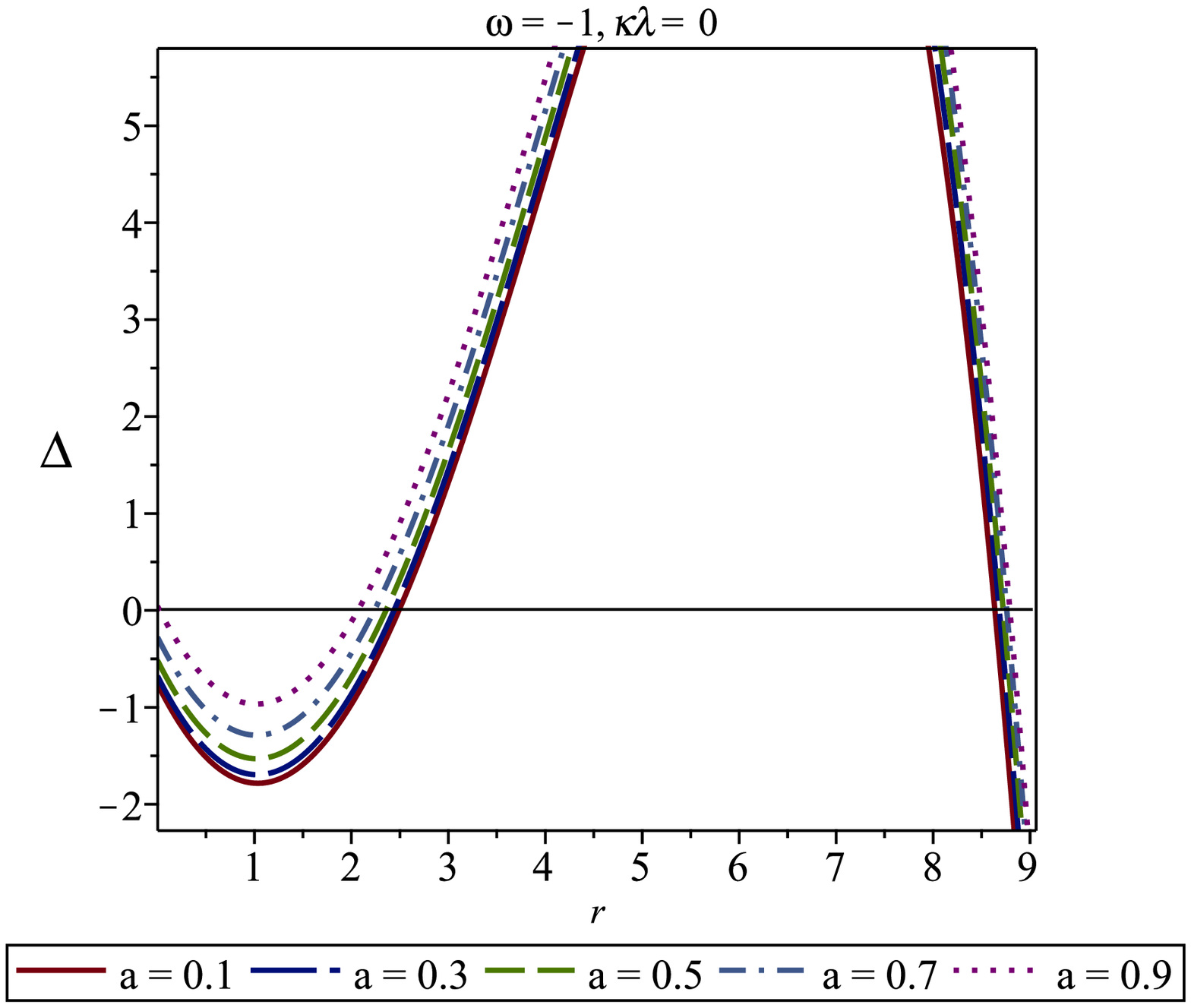}
		\put(40,20){\includegraphics[scale=0.11]{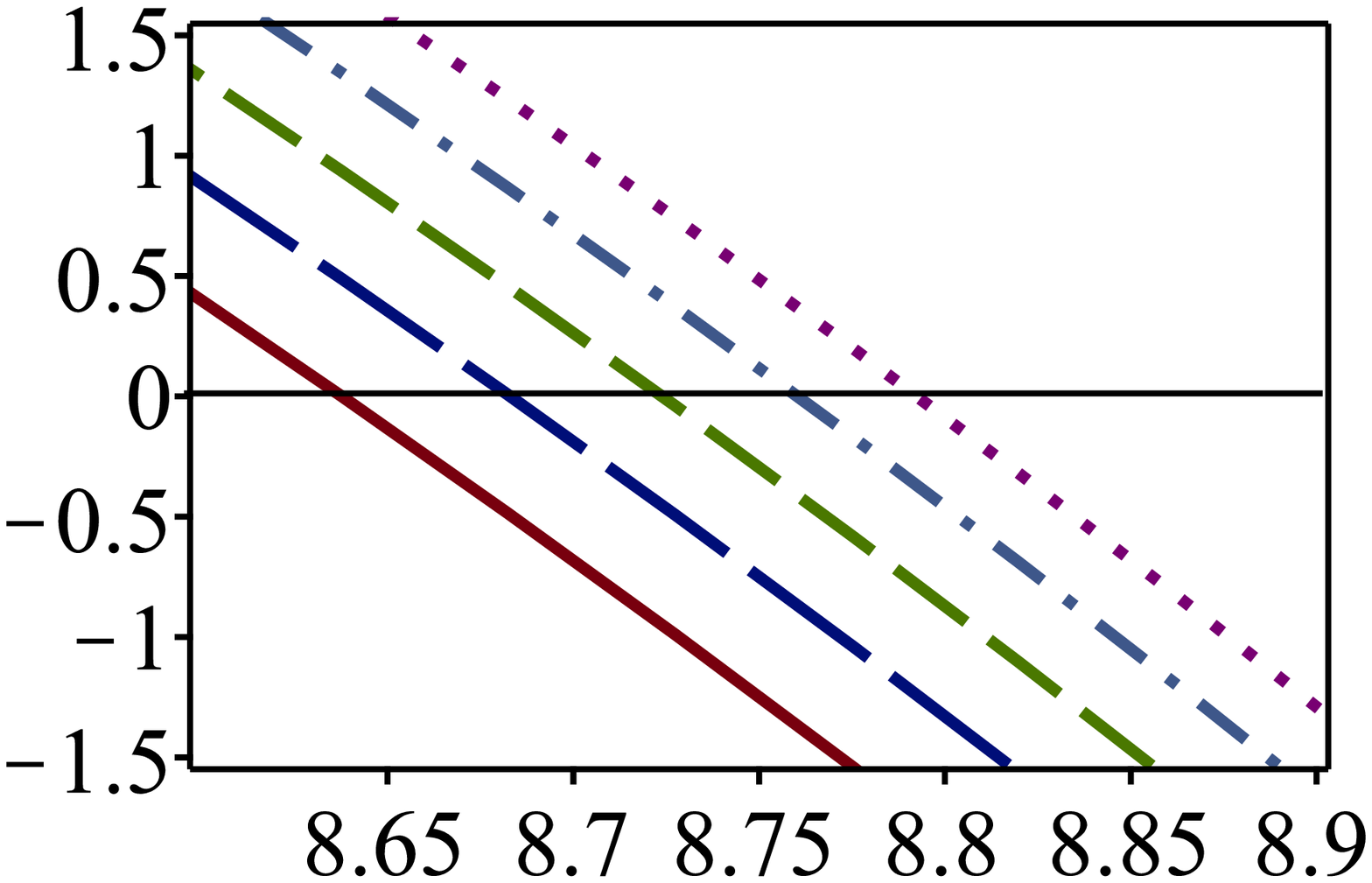}}
	\end{overpic}
	\includegraphics[scale=0.3]{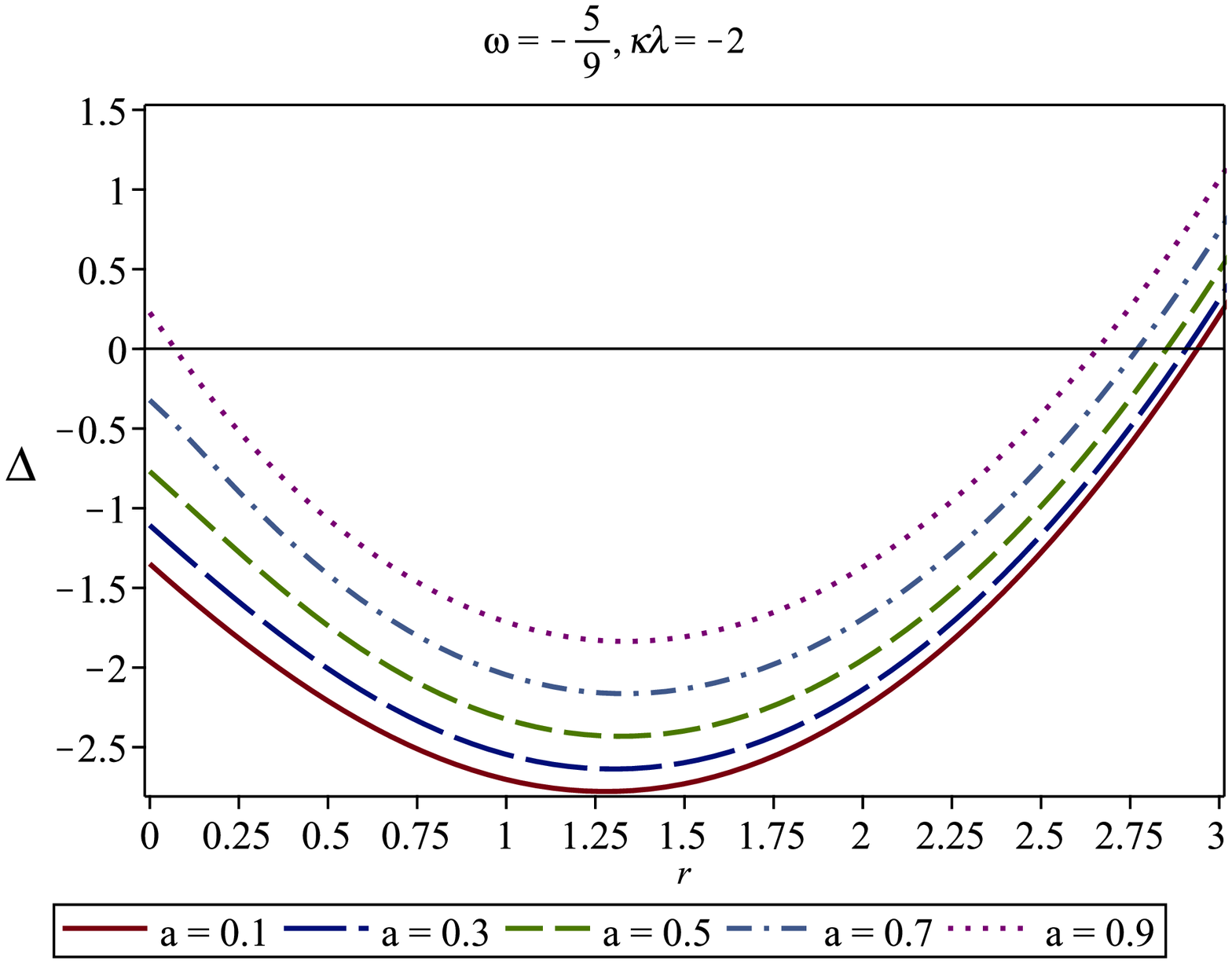}
	\caption{\label{gambar:1} Plot of $\Delta$ function with respect to coordinate $r$ at the north pole ($\theta=0$). Here we set $M=1,Q=0.5,N_s=1,n=0.88$ and vary the rotation parameter $a$. We have the solution with cosmological horizon $r_q$ (left) and the one that does not have cosmological horizon $r_q$ (right).}
\end{figure}

\begin{figure}
	\centering
	\includegraphics[scale=0.3]{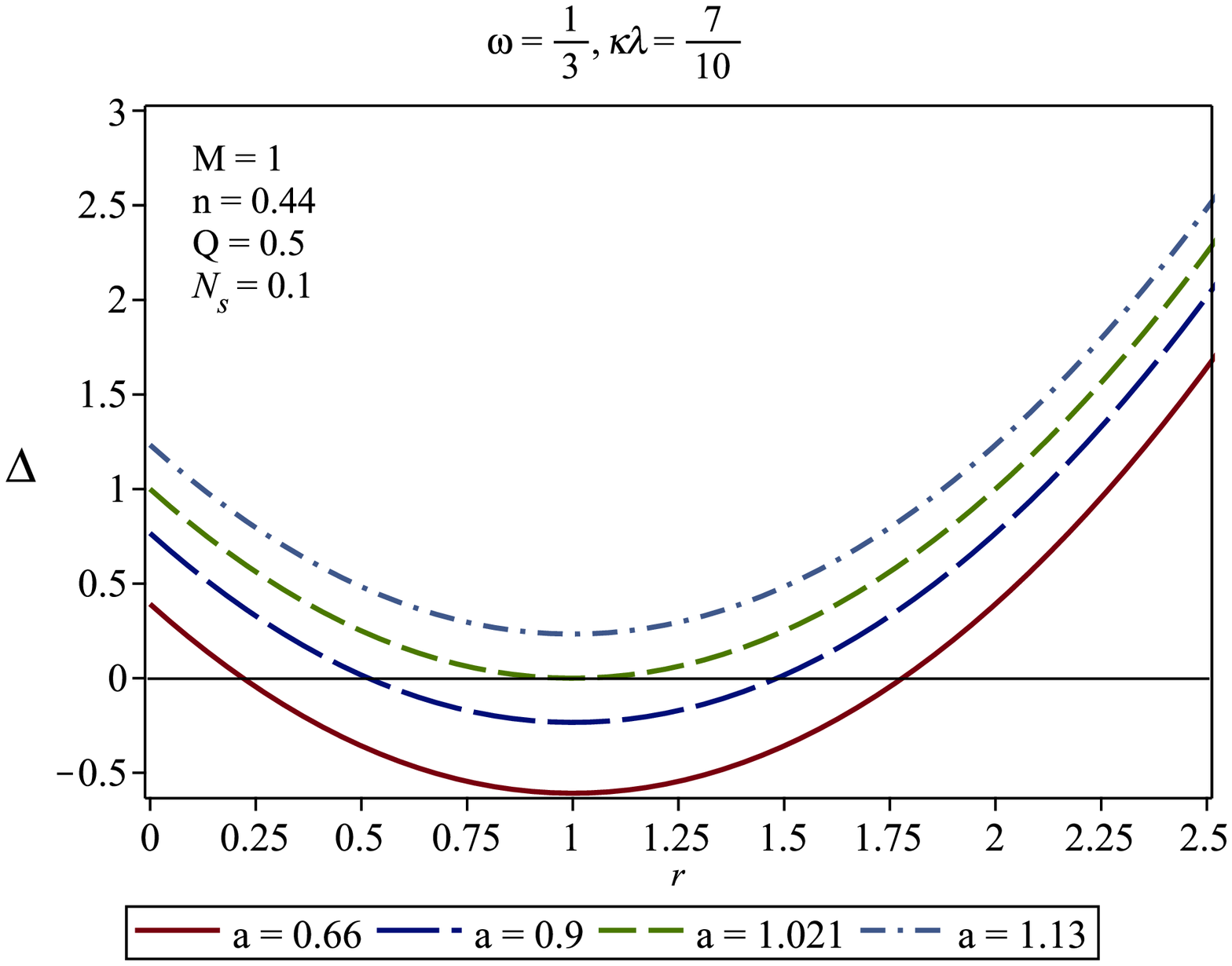}
	\includegraphics[scale=0.3]{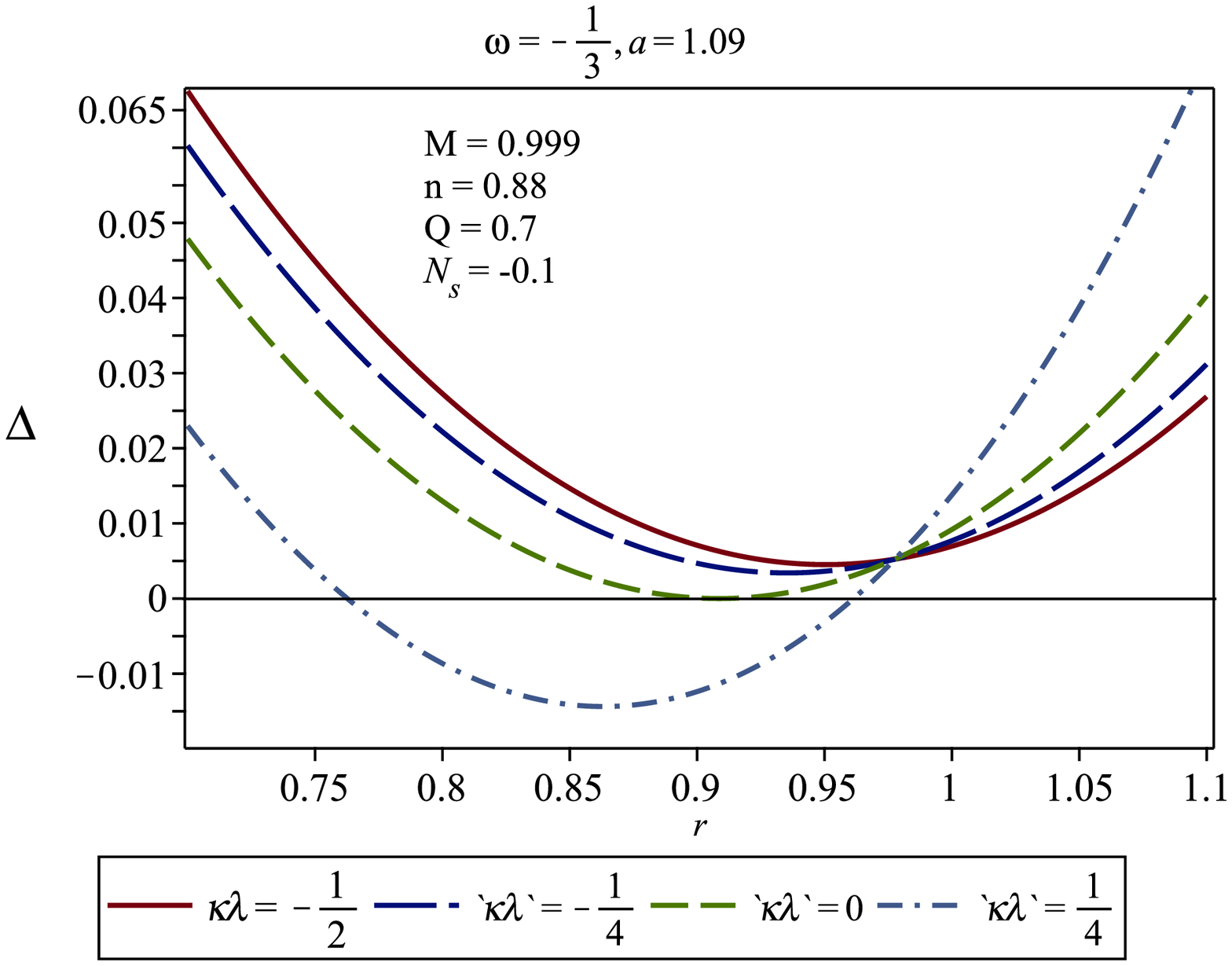}
	\caption{\label{gambar:2} Plot showing the variation of $\Delta$ function with respect to coordinate $r$ at $\theta=0$. In this figure we represent the extremal solution, \textit{i.e.} the solution with only one horizon. The extremal black hole solution is given by the \textit{green-dashed} line on both images.}
\end{figure}

Black hole horizon is the null surface that determined by the coordinate singularity $\Delta=0$. Since $\Delta$ is a function of $r$ and $\theta$, the $r$ solution is $\theta$ dependence. Thus, the KNN-R black hole horizons are determined by zeros of
\begin{equation}
r^2+a^2-n^2-2Mr+Q^2-N_s\Sigma^{\frac{2-\zeta}{2}}=0. \label{eq:42}
\end{equation}
From here we expect to have the inner ($r_-$), outer ($r_+$), and even cosmological horizon ($r_q$), if any, with $r_-<r_+<r_q$. Analytical expressions for the horizon solution are given by Table \ref{tabel:1}. It shows all possibilities for a quadratic solution. The value of $\omega$ is chosen specifically to match the common surrounding matters such as: dust field ($\omega=0$), radiation field ($\omega=1/3$), and quintessence field ($\omega=-1/3,-2/3$) \cite{25}. Hence the Rastall parameter has been chosen arbitrarily to match the solutions.
\begin{table}[H]
	\tbl{\label{tabel:1}Some possibilities of $\omega$ and $\kappa\lambda$ values that yield to the quadratic analytical horizon. The "+" and "-" sign indicates outer and inner horizons, respectively.}
	{\begin{tabular}{@{}cccc@{}} \toprule
			$\omega$,$\kappa\lambda$ &Horizon ($r_{\pm}$)\\\colrule
			$0$, $1/6$&$\frac{M}{(1-N_s)}\pm\frac{\sqrt{M^2-(1-N_s)(a^2-n^2+Q^2-N_s(a\cos\theta-n)^2)}}{(1-N_s)}$\\
			$1/3$, any&$M\pm\sqrt{M^2-(a^2-n^2+Q^2-N_s)}$\\
			$-1/3$, $0$&$\frac{M}{(1-N_s)}\pm\frac{\sqrt{M^2-(1-N_s)(a^2-n^2+Q^2-N_s(a\cos\theta-n)^2)}}{(1-N_s)}$\\ \botrule
		\end{tabular}}
\end{table}
\indent Furthermore, we are able to see the horizon solution aside from the analytical expression. We want to solve Eq. (\ref{eq:42}) numerically so we can use the parameters rather freely. Here we presents inner, outer, and cosmological horizon solutions for arbitrary parameter in Table \ref{tabel:2}, where $\omega=-\frac{2}{3}$ and $\omega=-1$ denotes the surrounding field for quintessence and cosmological constant, respectively \cite{25}. It shows that some suitable parameters are able to wipe out inner horizon. We can see a detailed behavior of the $\Delta$ function with respect to $r$ in Fig. \ref{gambar:1} and Fig. \ref{gambar:2}. In Fig. \ref{gambar:1} (left), we can see that the cosmological horizon $r_q$ is increasing, but the outer horizon is decreasing as the rotation parameter $a$ increases. On the different case without cosmological horizon (right), we also get the same behavior of the outer horizon. In Fig. \ref{gambar:2}, we have the solutions for extremal and non-extremal black hole. We also have curves that does not intersect with $\Delta=0$ line, and that represents the naked singularity since they don't have any horizons. Black holes with a naked singularity is prohibited by the Cosmic Censorship Conjecture, so it is unphysical.
\begin{table}[H]
	\tbl{\label{tabel:2}Table showing horizon solutions for $M=1,N_s=0.01,Q=0.09$ and for various value of $\omega,\kappa\lambda,a,$ and $n$ at the north pole ($\theta=0$). The "$-$" sign indicates that there is no possible solution for $r_-$.}
	{\begin{tabular}{@{}cccccccc@{}} \toprule
			$\omega,\kappa\lambda$ &$a$ &$n$ &$r_-$ &$r_+$ &$r_q$\\\colrule
			$-\frac{2}{3},0$&0.5&0.5&0.00405&2.03754&97.95840\\
			&&1&-&2.37276&97.94650\\
			&0.9&0.5&0.34170&1.69630&97.96185\\
			&&1&-&2.13092&97.95622\\
			$0,\frac{1}{4}$&0.5&0.5&0.00405&2.08702&8.78955\\
			&&1&-&2.46321&8.68901\\
			&0.9&0.5&0.34222&1.72871&8.81617\\
			&&1&-&2.18835&8.77171\\
			$-1,\frac{1}{3}$&0.5&0.5&0.00405&2.08702&8.78955\\
			&&1&-&2.46321&8.68901\\
			&0.9&0.5&0.34222&1.72871&8.81617\\
			&&1&-&2.18835&8.77171\\\botrule
	\end{tabular}}
\end{table}
\indent The behavior of KNN-R black hole horizon can gives us some important properties. As shown later in section 3 and 4, the thermodynamics and the particle geodesic of this black hole depends strongly on the horizon's behavior. Black holes with no cosmological horizon $r_q$ are easier to be investigated, it only depends on the outer horizon $r_+$.
\subsection{Ergoregion Structure}
\begin{figure}
	\begin{center}
		\includegraphics[scale=0.22]{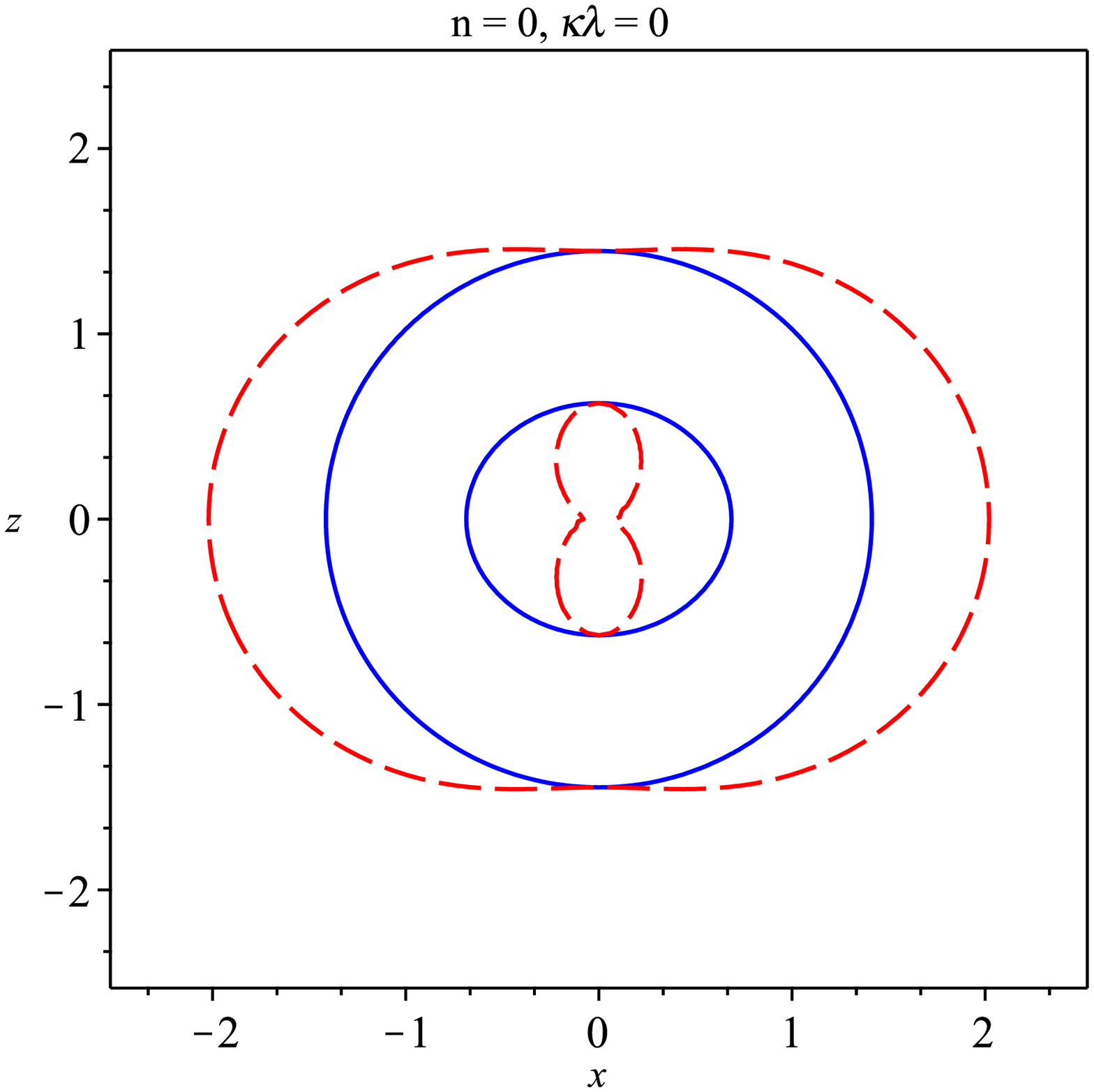}\includegraphics[scale=0.22]{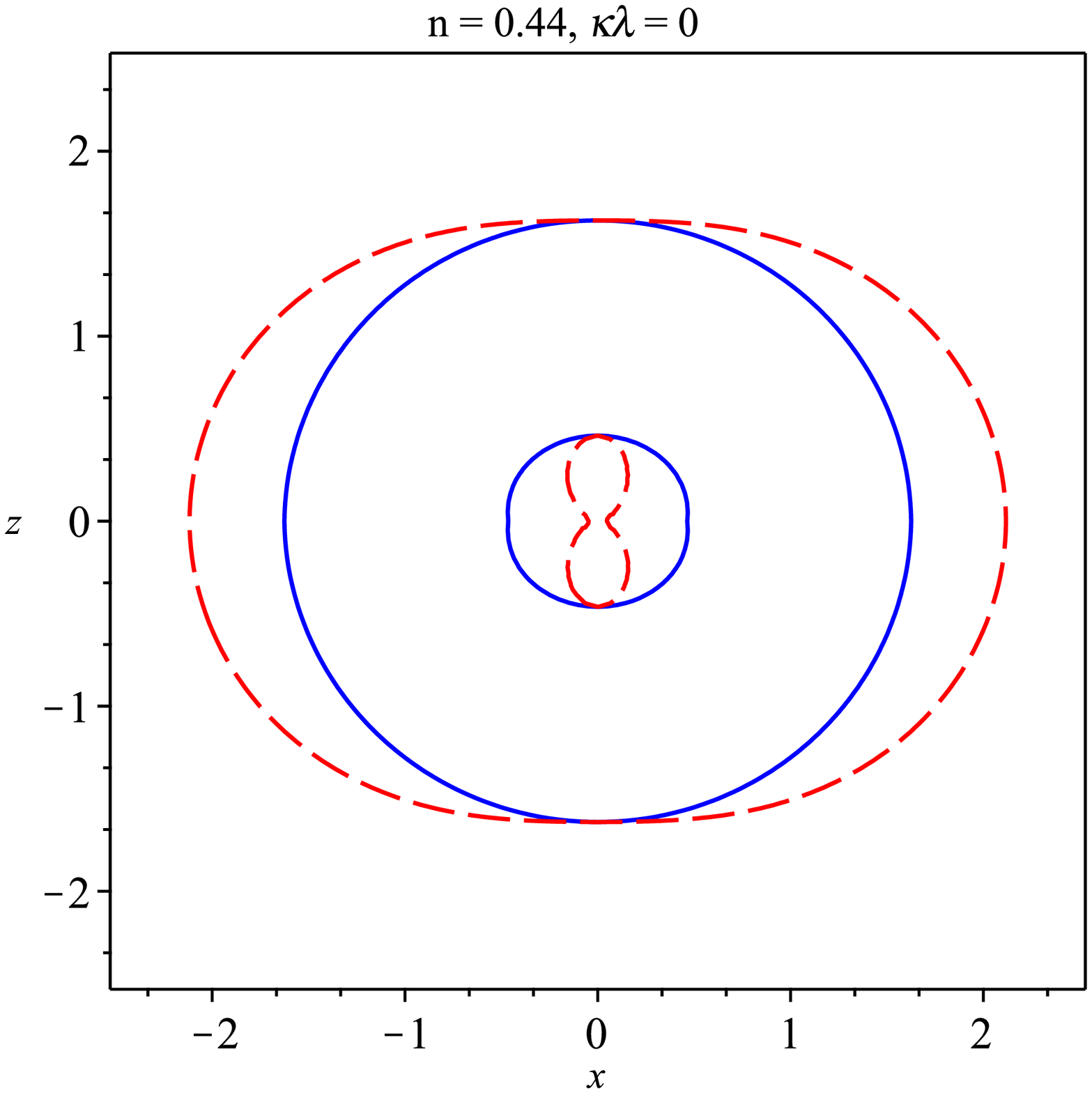}\includegraphics[scale=0.22]{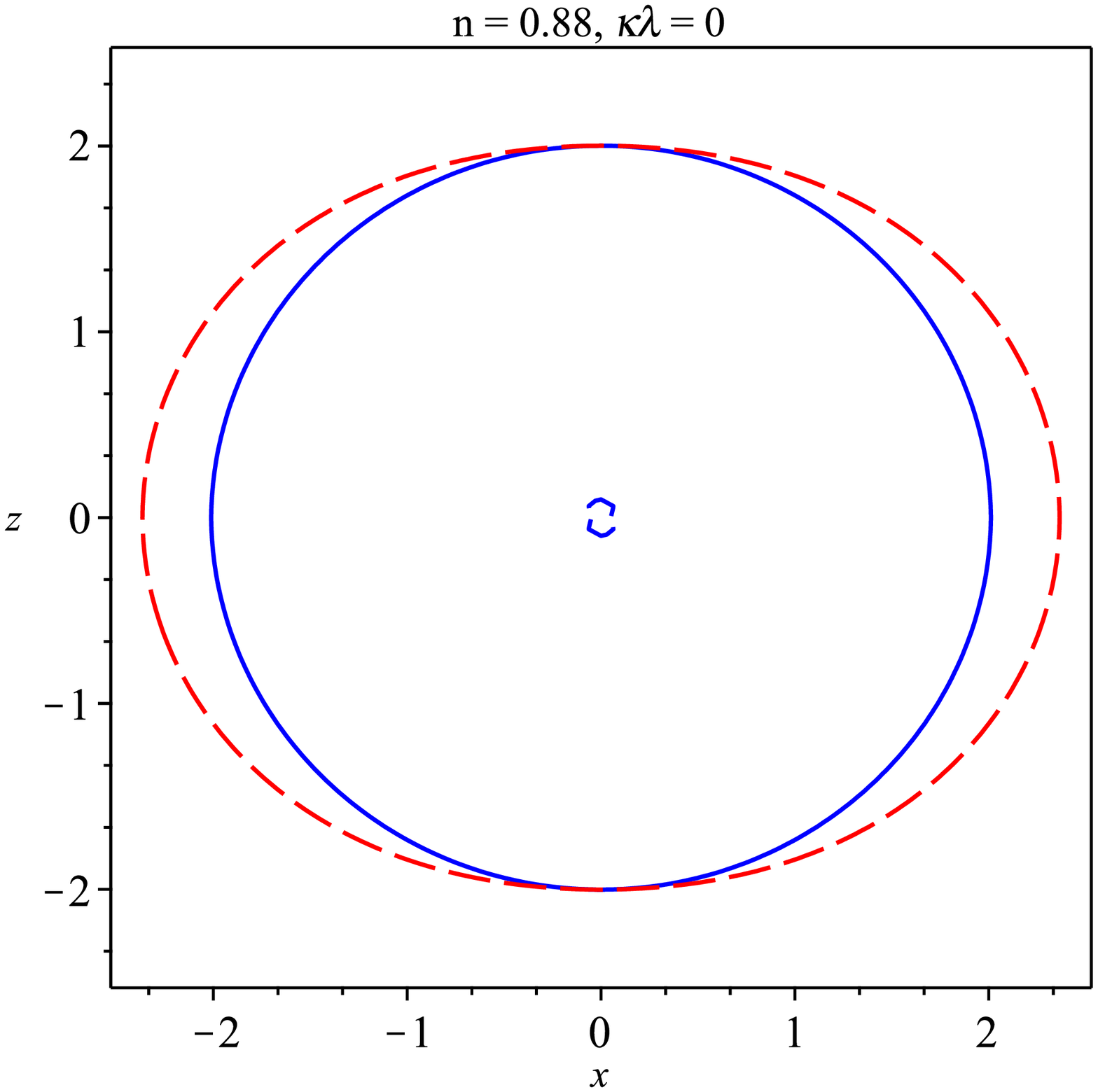}\\
		\includegraphics[scale=0.22]{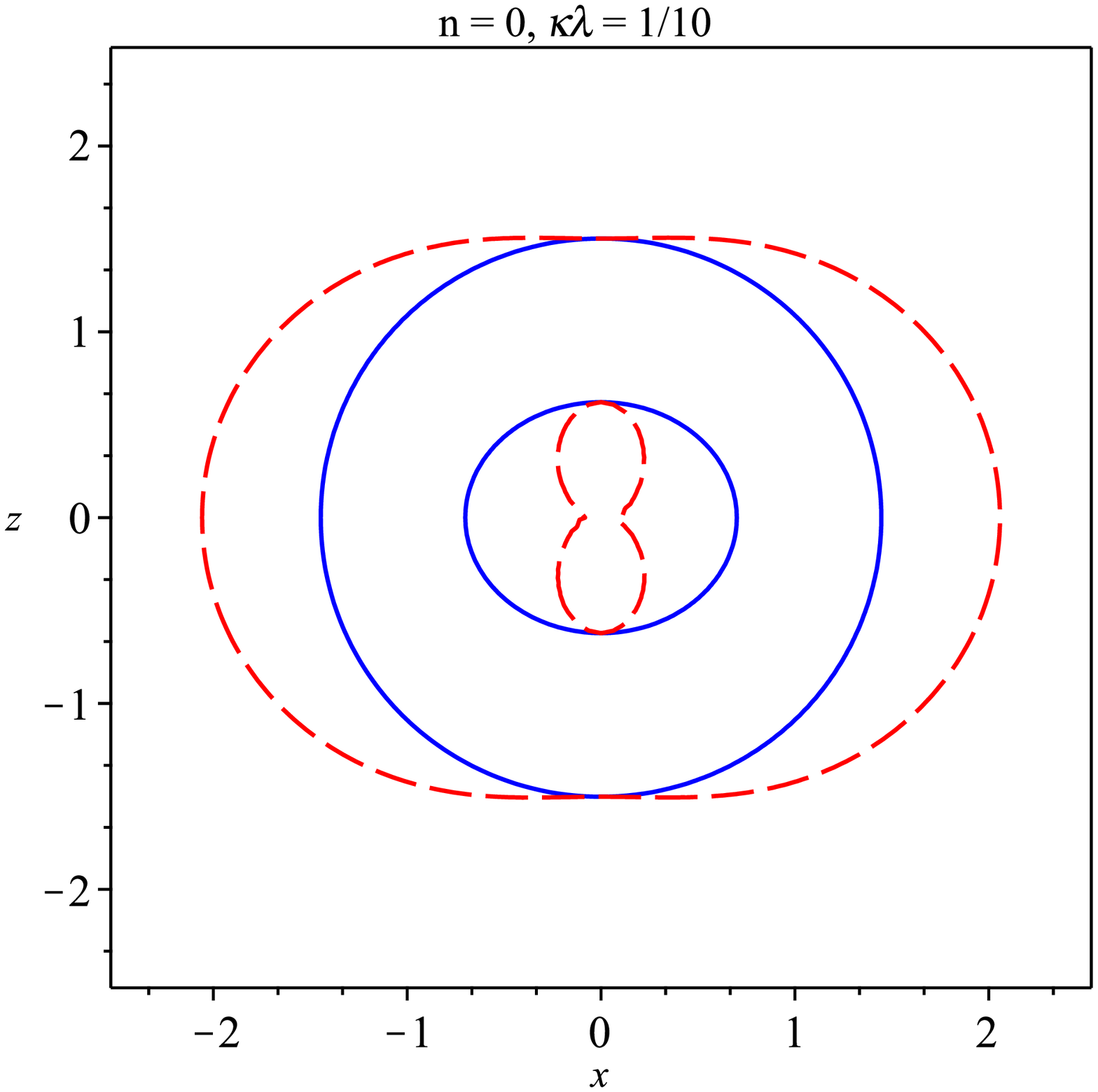}\includegraphics[scale=0.22]{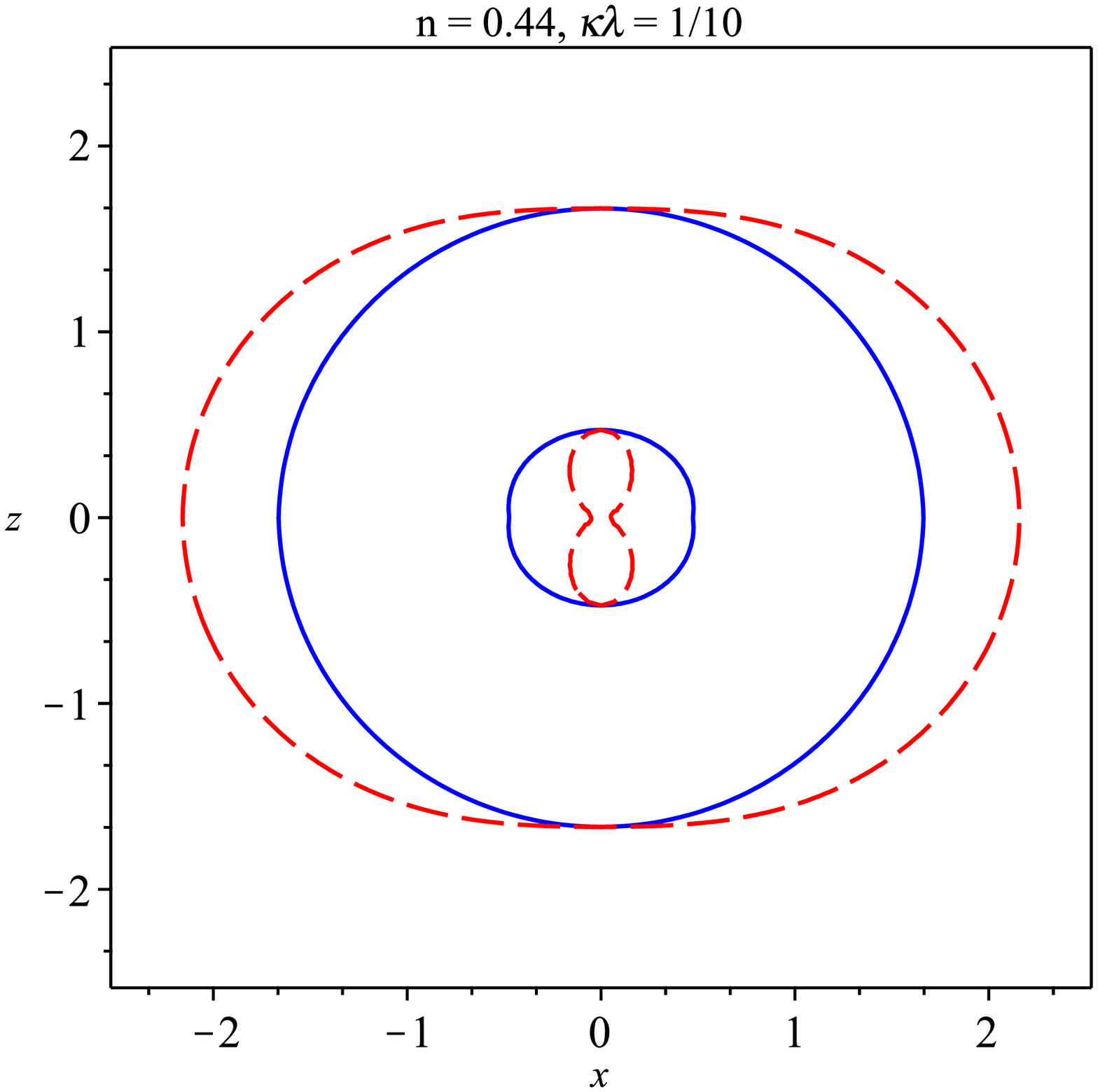}\includegraphics[scale=0.22]{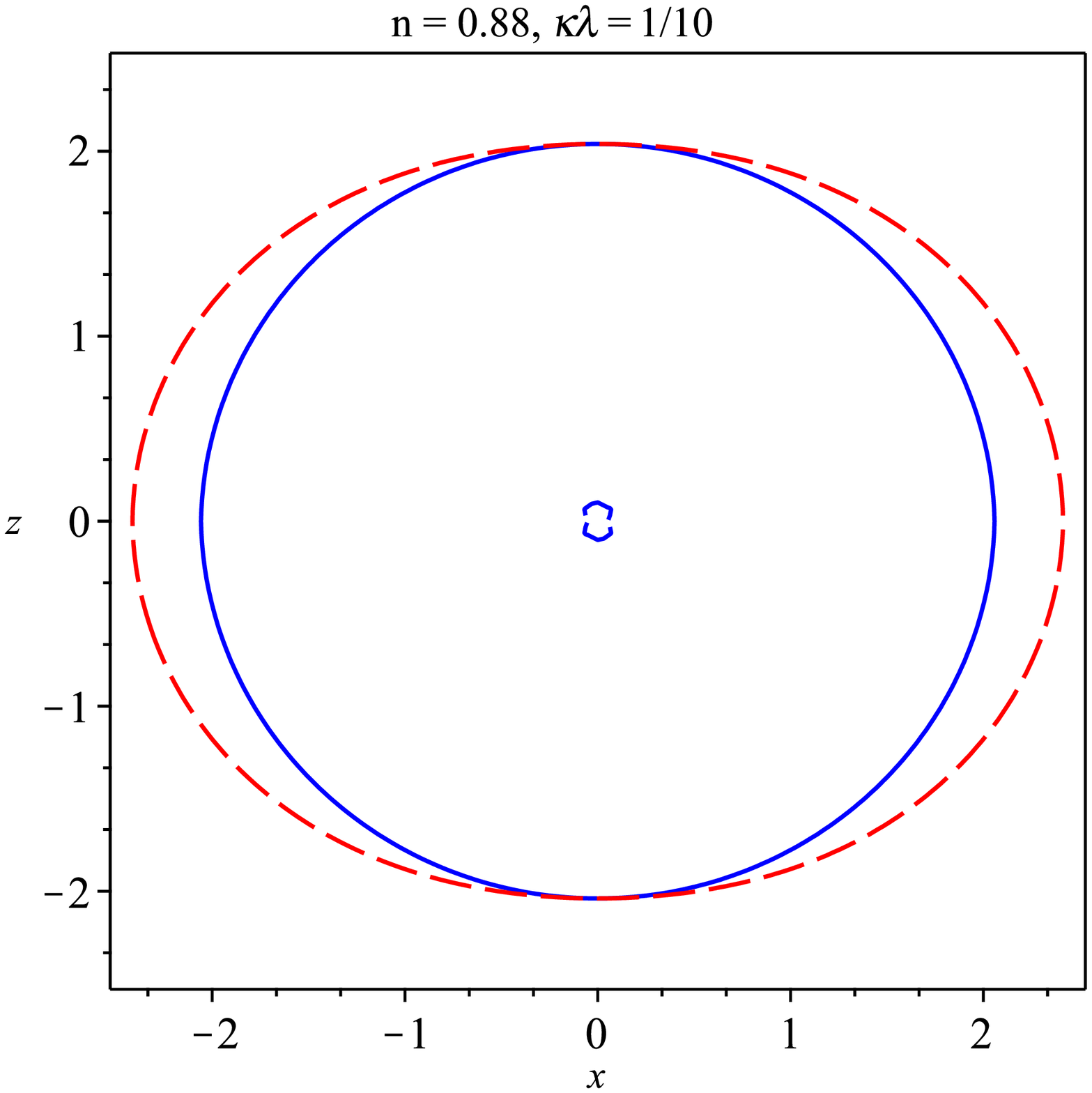}\\
		\includegraphics[scale=0.22]{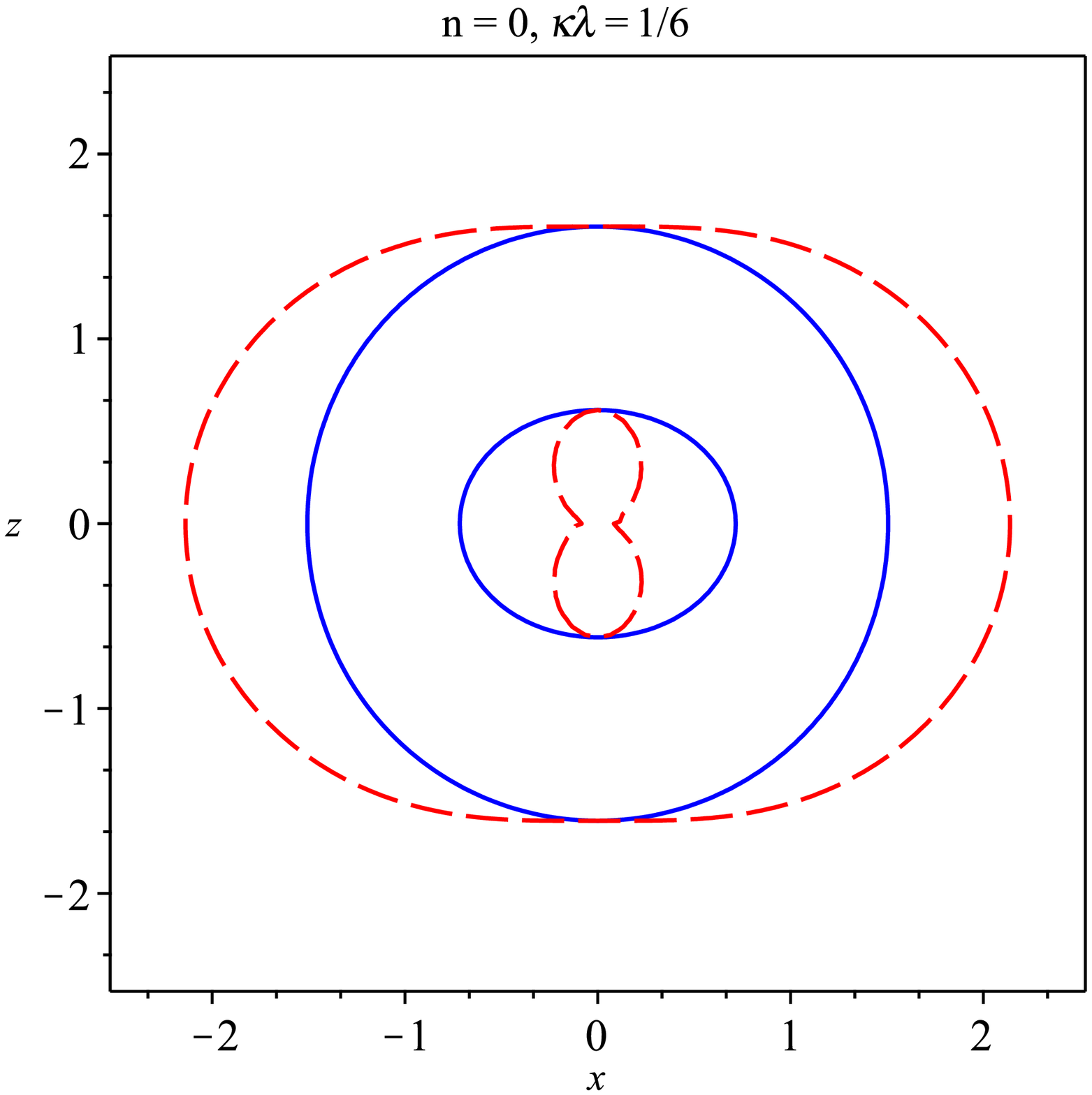}\includegraphics[scale=0.22]{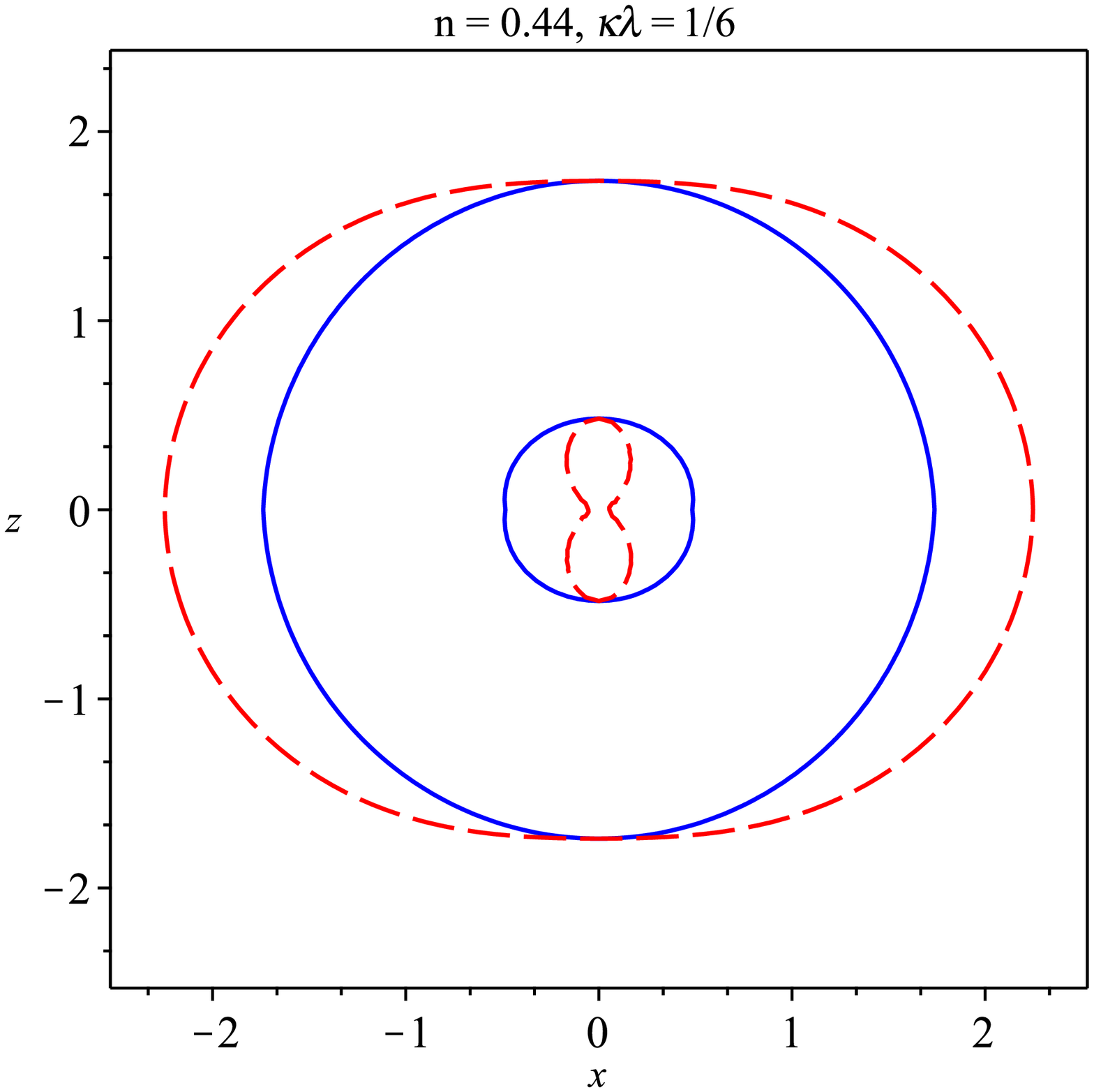}\includegraphics[scale=0.22]{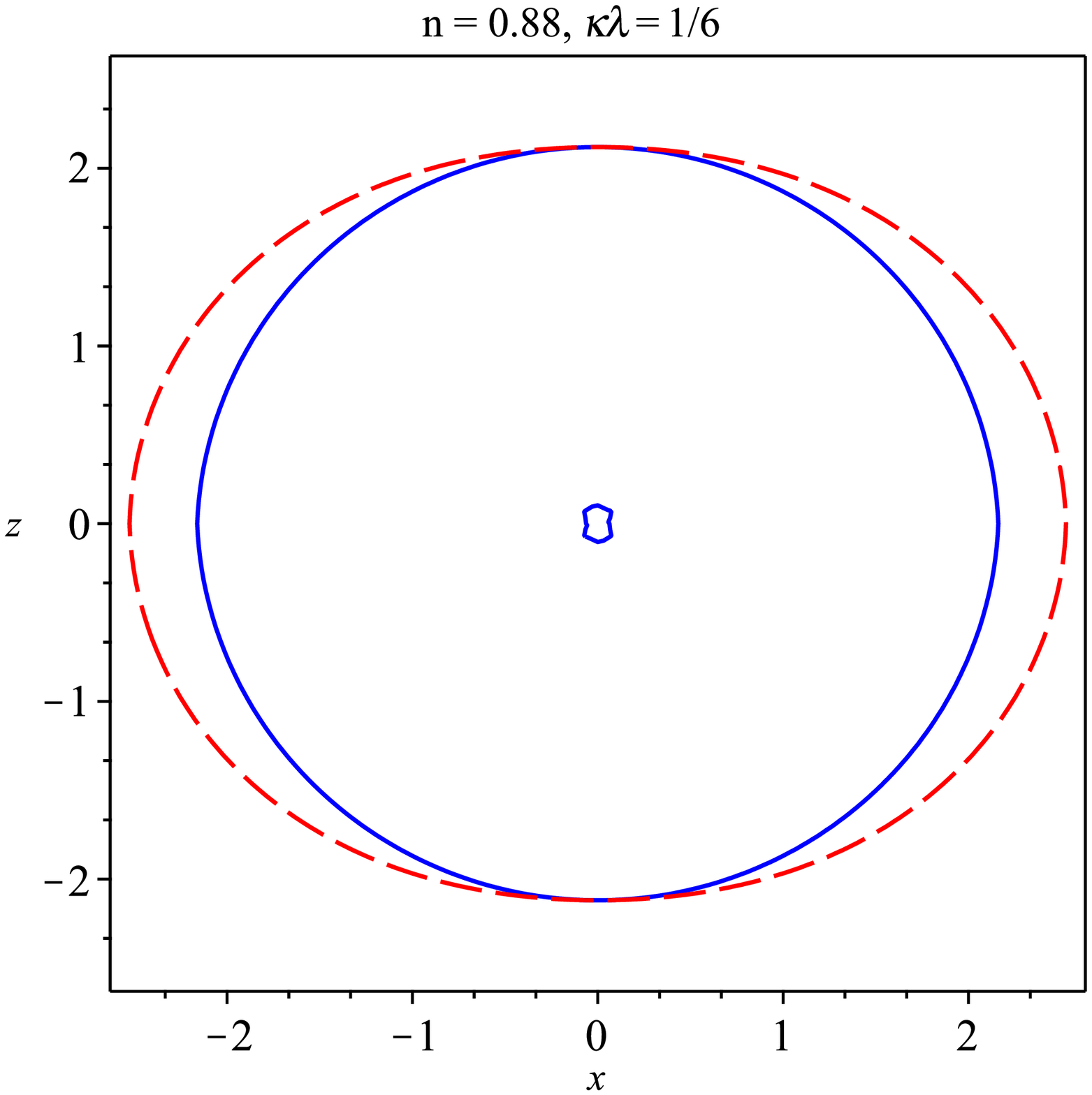}\\
		\includegraphics[scale=0.22]{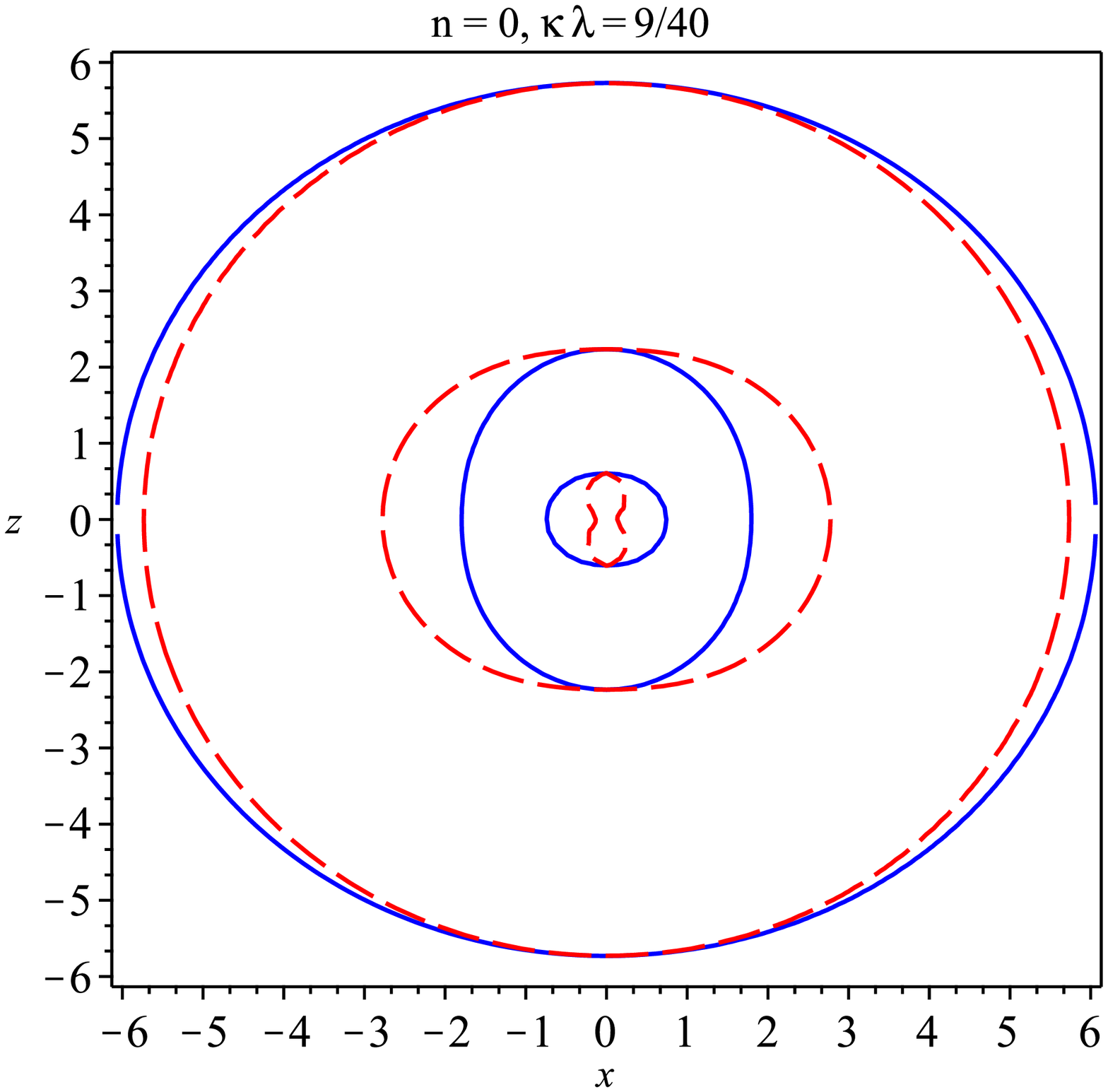}\includegraphics[scale=0.22]{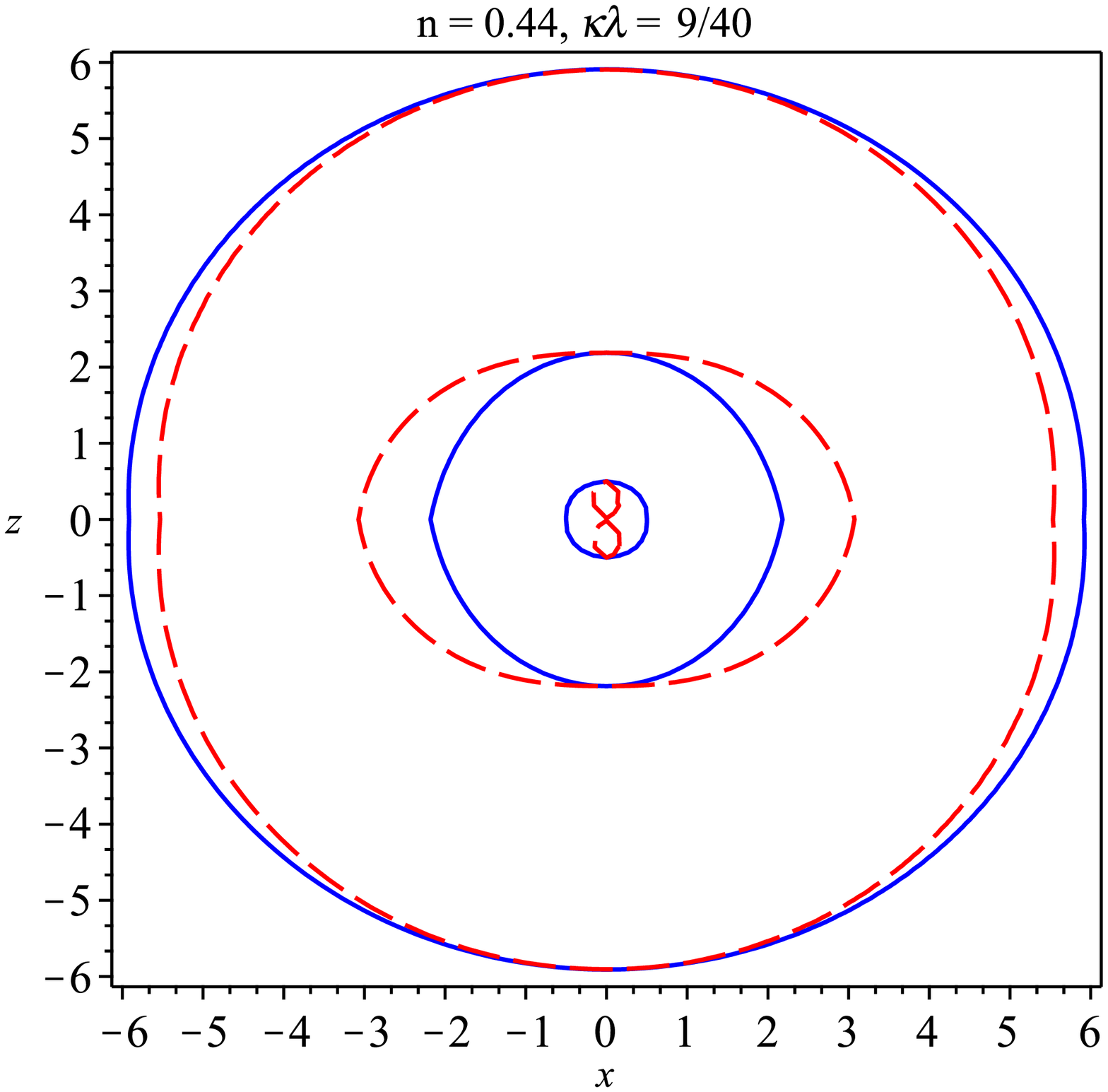}\includegraphics[scale=0.22]{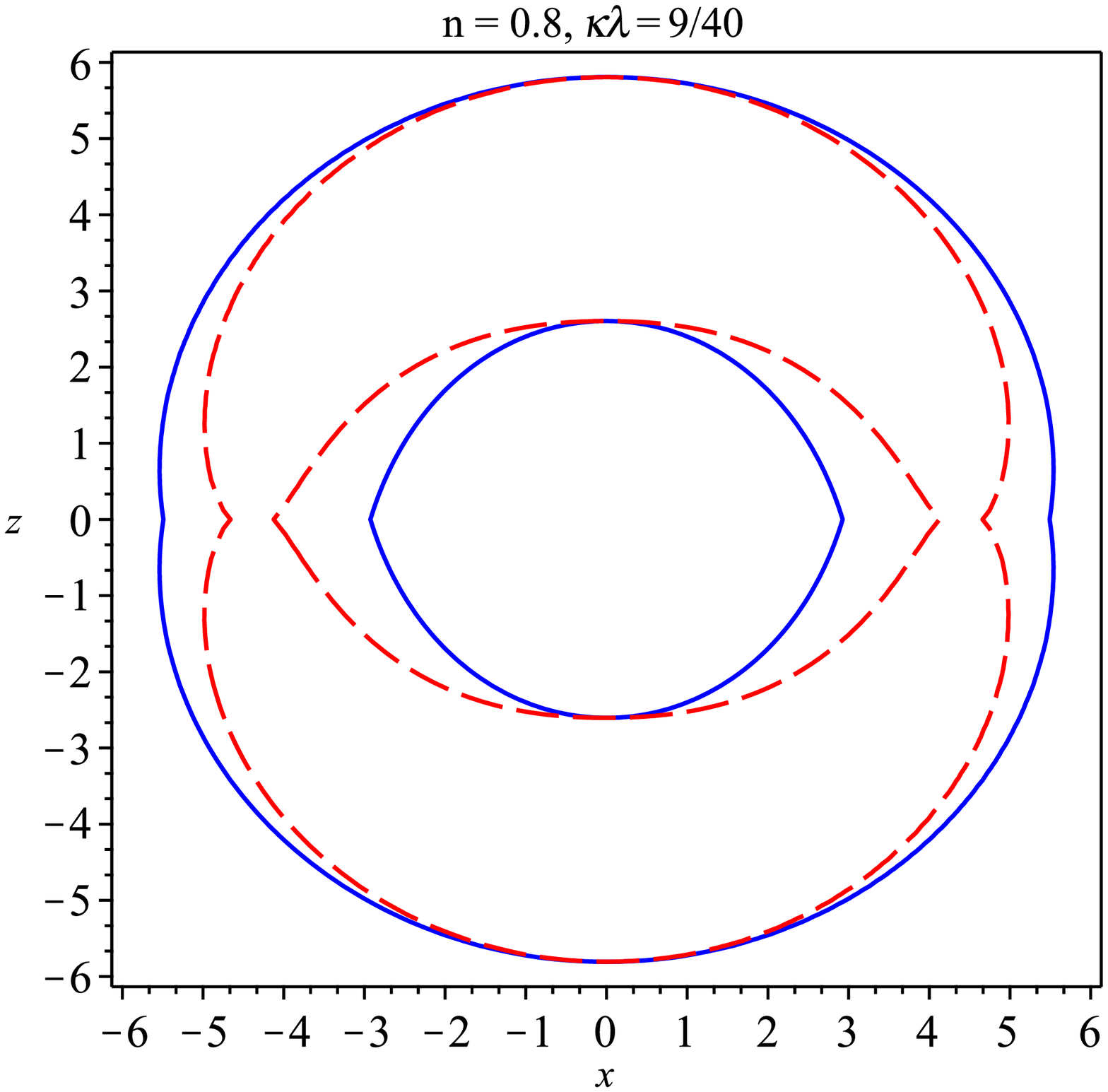}
		\caption{Plot showing the structure of the horizon and ergosurface with surrounding dust field $\omega=0$, $M=1$, $Q=0.4$, $a=0.9$, and $N_s=0.1$ with variations of NUT $n$ and Rastall parameters. The \textit{red-dashed} and \textit{blue-solid} line indicated the ergosurface and horizon, respectively. As the NUT parameter $n$ increases, $r_+$ is slightly bigger, but $r_-$ gets smaller. However, we discover the existence of $r_q$ when $\kappa\lambda=1/4$.}\label{gambar:3}
	\end{center}
\end{figure}

An axisymmetric black hole, such as this one, will have a Killing vector in the form of $K^{\mu}=\xi^{\mu}_{(t)}+\Omega_H\xi^{\mu}_{(\phi)}$, where $\Omega_H$ is the angular velocity of the zero angular momentum observer (ZAMO) that will be investigated further, $\xi^{\mu}_{(t)}$ and $\xi^{\mu}_{(\phi)}$ are vectors associated with time and rotational translation invariance, respectively. A null time-like hypersurface is then given by $\xi^{\mu}_{(t)}\xi_{\mu(t)}=0$, which leads to 
\begin{equation}
g_{tt}=-\bigg(\frac{\Delta-a^2\sin^2\theta}{\Sigma}\bigg)=0,
\end{equation}
as the static limit surface and forms the ergosurface of the black hole, \textit{i.e.} the boundary between ergoregion and an asymptotically flat time-like region. Therefore, the $r$ solution for
\begin{equation}
r^2+a^2\cos^2\theta-n^2-2Mr+Q^2-N_s\Sigma^{\frac{2-\zeta}{2}}=0 \label{eq:43},
\end{equation}
gives the ergosurface. Even though it also depends on coordinate $\theta$, this surface differs from the horizon and it coincides at the north and south pole. Equation (\ref{eq:43}) can also gives us more than one solutions. \\
\indent Since previously we have been working only on equatorial case, now we represent how the $\theta$ dependence act on the horizon and ergoregion structure, given by Fig. \ref{gambar:3}. We can see from the figure that the horizon size of a KNN-R black hole surrounded by dust field is raising along with the increasing of NUT parameter. The inner horizon of the black hole is barely noticeable. We also have the cosmological horizon at $\kappa\lambda=9/40$ (here we take $9/40$ instead of $1/4$ to avoid naked singularity). The cosmological horizon $r_q$ is getting smaller when the NUT parameter is increasing. It is possible to have an ergosphere radius smaller than the cosmological horizon.
\subsection{Zero Angular Momentum Observer}
Inside the ergosphere, we can have an observer that seems to be rotating around the black hole seen by a distant observer but with vanishing angular momentum, called the Zero Angular Momentum Observer or ZAMO for short. The angular velocity of ZAMO is given by \cite{52}
\begin{equation}
\Omega=-\frac{g_{t\phi}}{g_{\phi\phi}}.
\end{equation}
Using the KNN-R line element (\ref{eq:40}), we obtain that
\begin{equation}
\Omega=\frac{a(\Sigma+a\chi)\sin^2\theta-\chi\Delta}{(\Sigma+a\chi)^2\sin^2\theta-\chi^2\Delta},
\end{equation}
and the angular velocity of ZAMO at the horizon is
\begin{equation}
\Omega_H=\frac{a}{r^2_h+n^2+a^2},
\end{equation}
where $r_h$ can be an inner, outer, or cosmological horizon. In many cases, the outer horizon is used to determine the angular velocity of a black hole. Unlike Newtonian mechanics, here we can have non-vanishing angular velocity even though the angular momentum is zero. This can happen because in General Relativity, the angular momentum corresponds to the space-time geometry \cite{52}. This phenomena also often called the \textit{dragging frame} effect \cite{25}.\\
\indent We can also write the KNN-R black hole line element (\ref{eq:40}) in terms of ZAMO as
\begin{align}
ds^2=&-\bigg[\frac{\Delta\Sigma\sin^2\theta}{(\Sigma+a\chi)^2\sin^2\theta-\chi^2\Delta}\bigg]dt^2+\frac{\Sigma}{\Delta}dr^2+\Sigma d\theta^2\nonumber\\
&+\frac{1}{\Sigma}((\Sigma+a\chi)^2\sin^2\theta-\chi^2\Delta)(d\phi-\Omega dt)^2.\label{eq:48}
\end{align}

\section{Thermodynamic Properties}
One of the most important discussions about black hole dynamics is the aspect of thermodynamics. Since the area law, \textit{i.e.} the law of non-decreasing area of a black holes, has been discovered, physicist formulated the classical thermodynamics of a black hole that connects the black hole parameters such as Mass, Spin, and Charge with the thermodynamical quantities. The entropy of a black hole, formulated by Bekenstein and Hawking, is represented by a quarter of the black hole horizon area \cite{53}.\\
\indent In this section, we only discuss a slowly rotating KNN-R black hole, and assume that $a<<n$ to get rid with the $\theta$ dependence of the horizon since in this limit, the $\Delta$ function is hardly depends on $\theta$. We want to avoid the $\theta$ dependence of the Hawking Temperature, \textit{i.e.} the temperature of the horizon, as well. This assumption is valid since in Fig. \ref{gambar:3}, the $\theta$ dependence of the horizon is also hardly apparent even for a not-really-slowly-rotating KNN-R black hole.
\subsection{Entropy and Surface Gravity}
The outer horizon area of KNN-R black hole, when the slowly-rotating assumption was applied, is given by
\begin{align}
\mathcal{A}_H&=\int_{0}^{2\pi}\int_{0}^{\pi}\sqrt{g_{\theta\theta}g_{\phi\phi}}\;d\theta d\phi\bigg|_{r=r_+}\nonumber\\
&=4\pi(r_+^2+a^2+n^2)\approx4\pi(r_+^2+n^2).\label{eq:49'}
\end{align}
The area forms a spherical shape with a radius, say, $r_{eff}^2=r_+^2+a^2+n^2\approx r_+^2+n^2$. Here, $r_+$ is just the outer horizon of the KNN-R black hole, without having to know the explicit formula analytically. Following the Bekenstein-Hawking formula for the entropy of a black hole \cite{53}, we may write (in Planck units)
\begin{align}
S_{\text{BH}}&=\frac{\mathcal{A}_H}{4}\nonumber\\
&=\pi(r_+^2+a^2+n^2)\approx\pi(r_+^2+n^2).\label{eq:entropi}
\end{align}
The dependence of the other KNN-R black hole parameters, such as $Q, N_s, \omega, \kappa\lambda$, on the Bekenstein-Hawking entropy is contained explicitly on the outer horizon $r_+$. This expression reduces to the Bekenstein-Hawking entropy for a Schwarzschild black hole when $a,n\rightarrow 0$.\\
\indent Surface gravity $\kappa$ is defined by\cite{25}
\begin{equation}
\kappa^2=-\frac{1}{2}\nabla^{\mu}K^{\nu}\nabla_{\mu}K_{\nu},
\end{equation}
where $K^{\mu}=\xi^{\mu}_t+\Omega_H\xi^{\mu}_{\phi}$ is a Killing vector for KNN-R space-time. A straight calculation gives the explicit form of the surface gravity for a slowly-rotating KNN-R black hole, as
\begin{equation}
\kappa=\frac{\Delta'(r_+)}{2(r_+^2+a^2+n^2)}\approx\frac{\Delta'(r_+)}{2(r_+^2+n^2)},\label{eq:52'}
\end{equation}
where prime notation denotes a derivative with respect to $r$, $\Delta'(r)\equiv d\Delta(r)/dr$. Equation (\ref{eq:52'}) vanishes in the extremal case since we will have $\Delta(r_+)=\Delta'(r_+)=0$.
\subsection{Temperature and Heat Capacity}
\begin{figure}
	\begin{center}
		\includegraphics[scale=0.3]{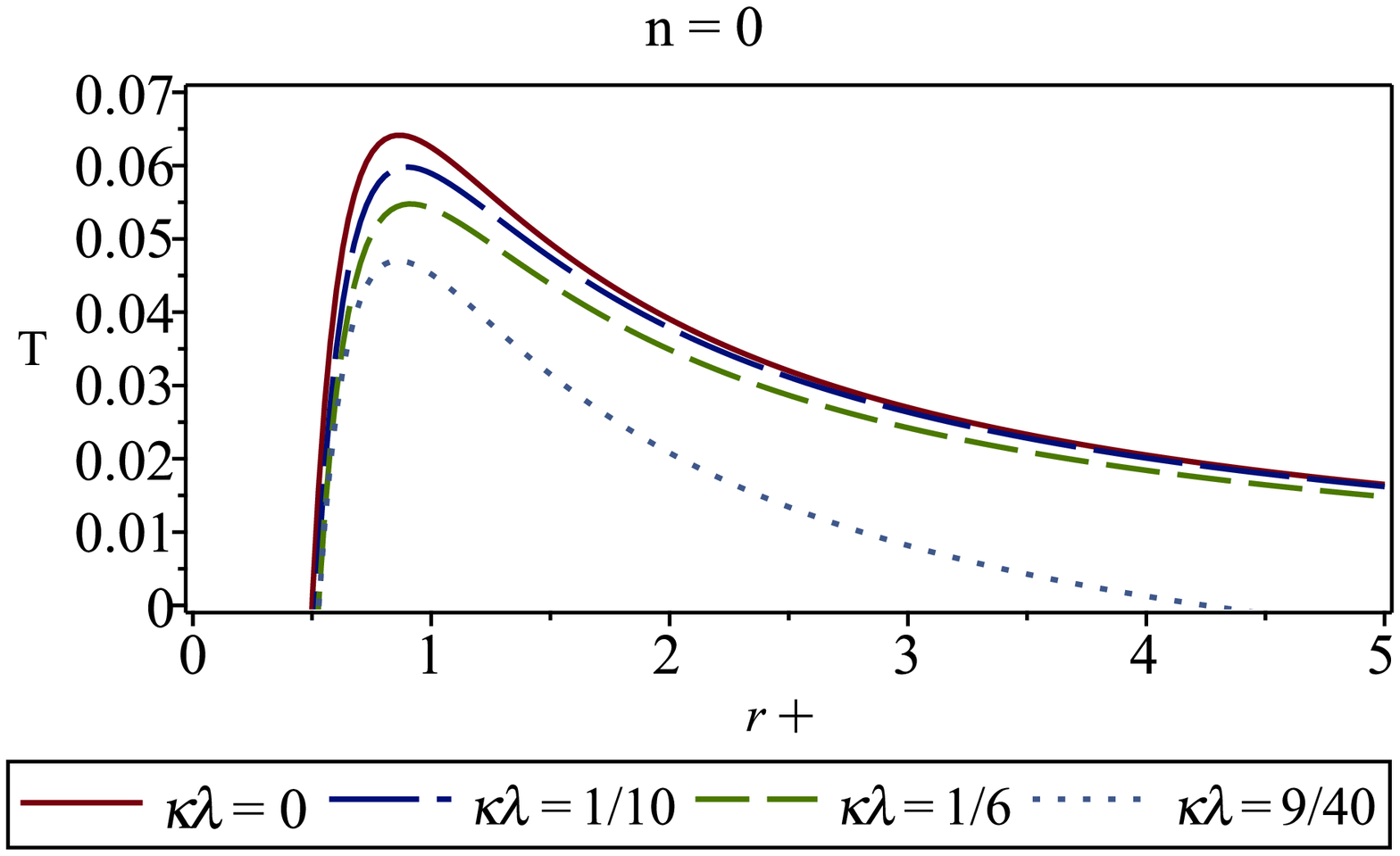}\includegraphics[scale=0.3]{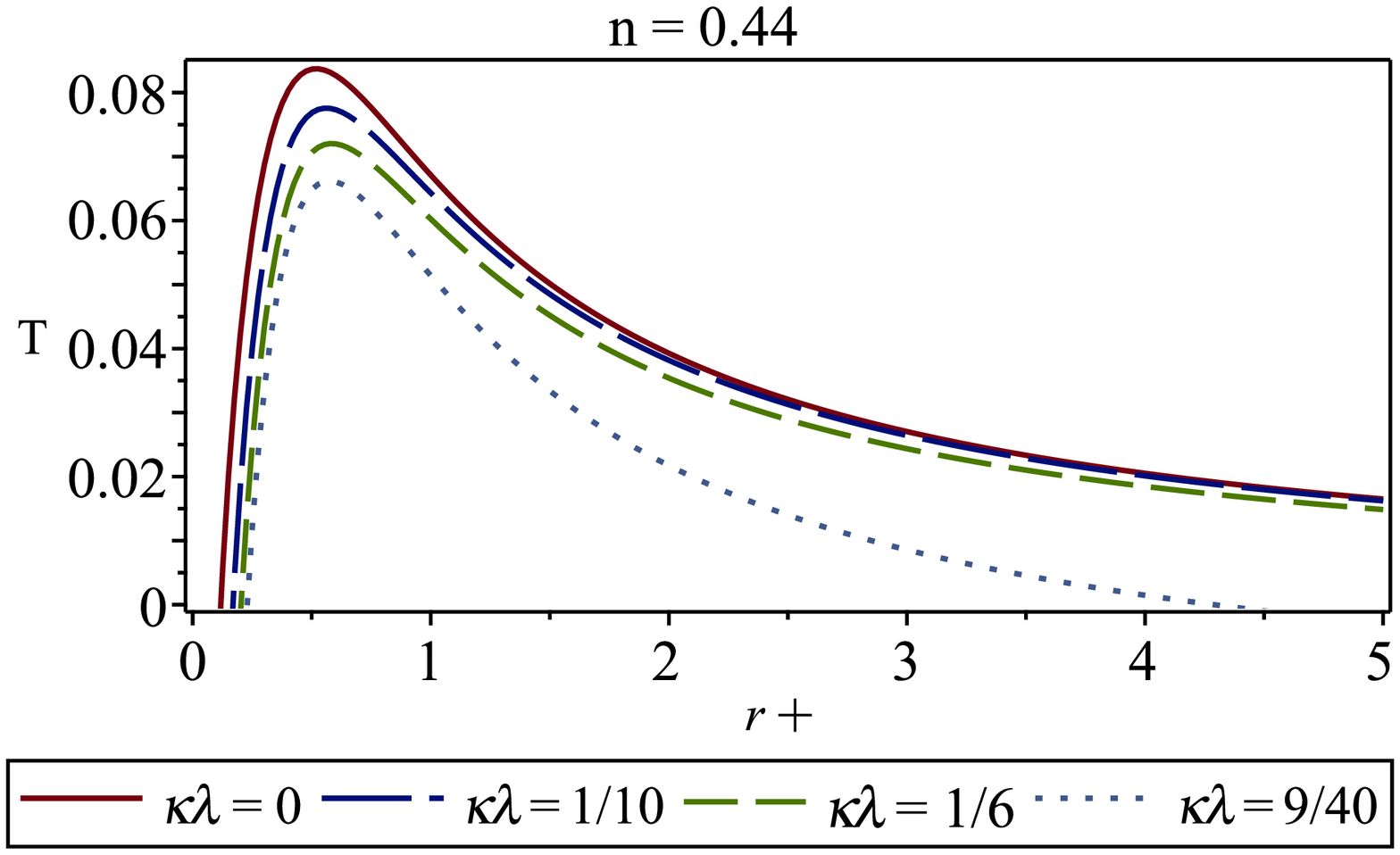}\\
		\includegraphics[scale=0.3]{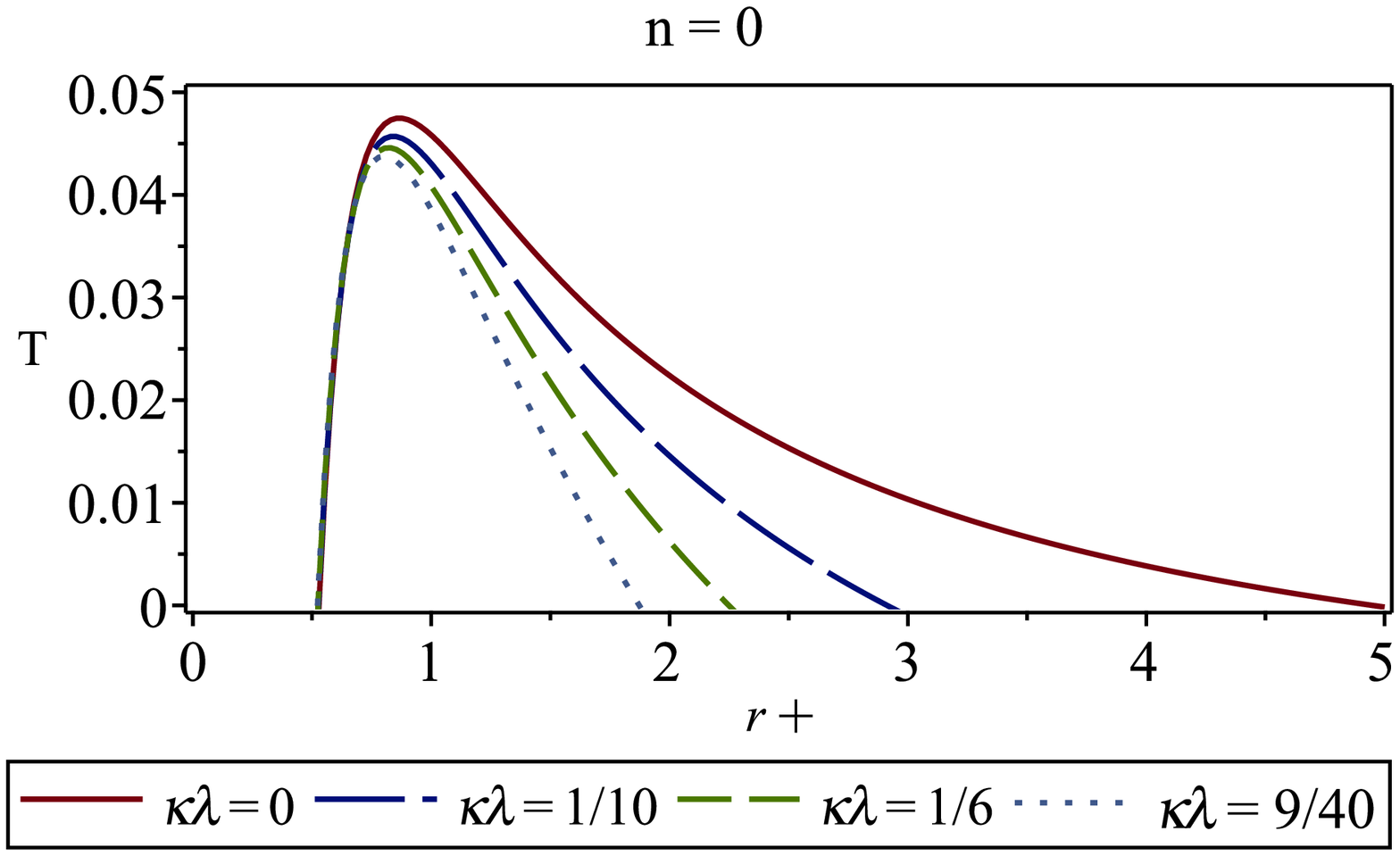}\includegraphics[scale=0.3]{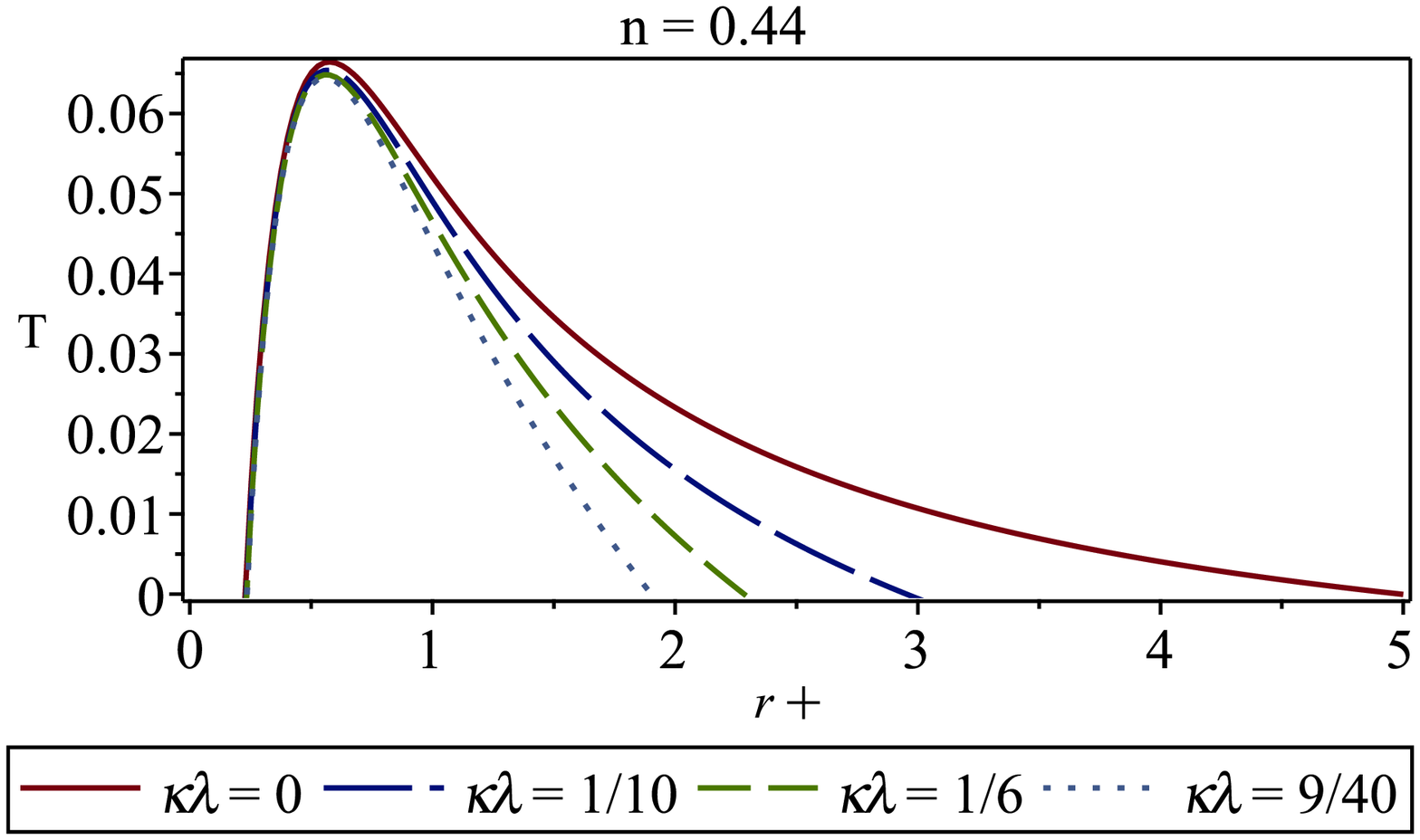}
		\caption{Plot showing the temperature of a KNN-R black hole horizon surrounded by dust field (top) and quintessence field (bottom) when the outer horizon $r_+$ varies. Here we use $Q=0.5$, $a\approx0$, and $N_s=0.1$.}\label{gambar:6}
	\end{center}
\end{figure}

\begin{figure}
	\begin{center}
		\includegraphics[scale=0.3]{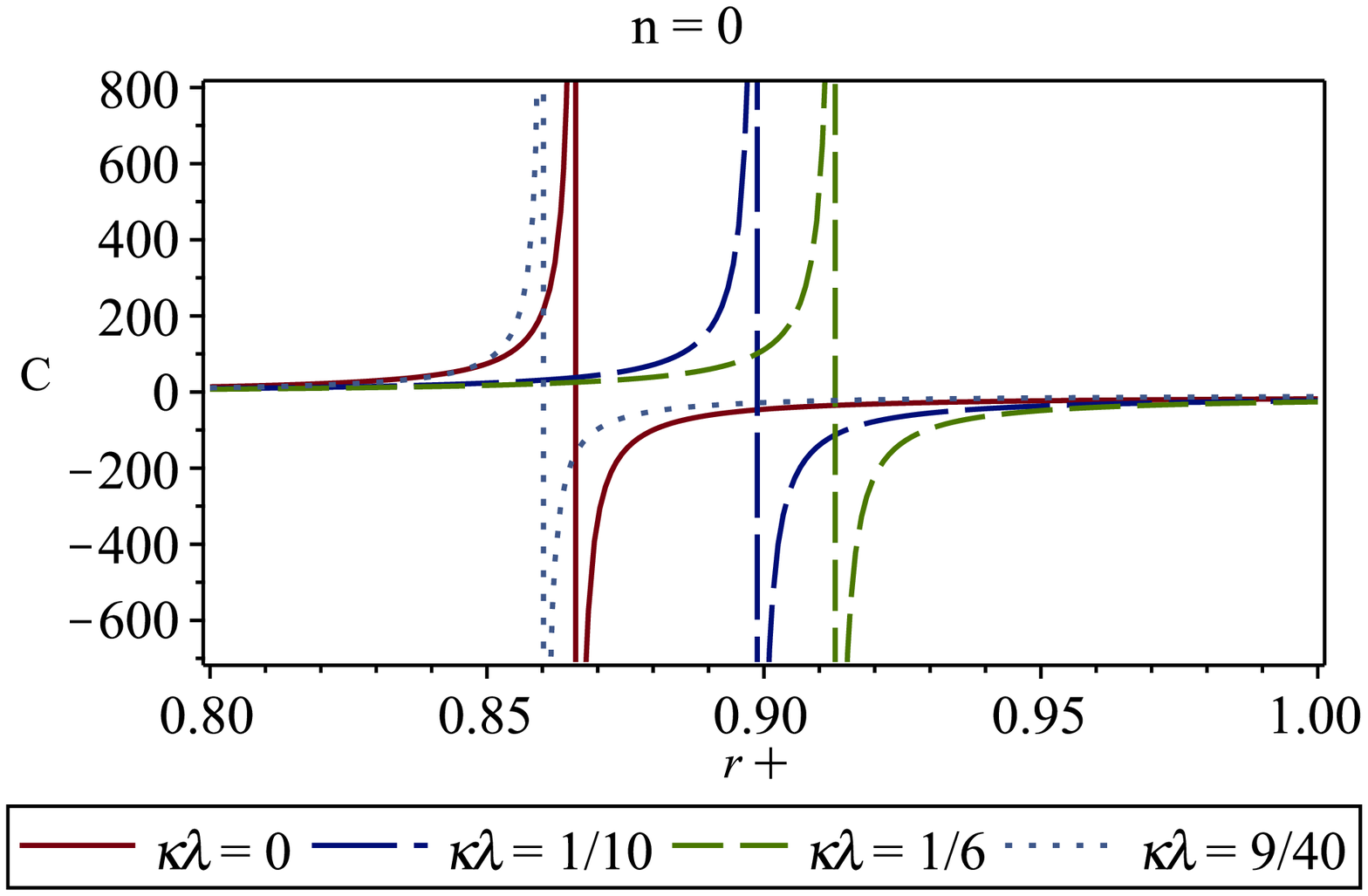}\includegraphics[scale=0.3]{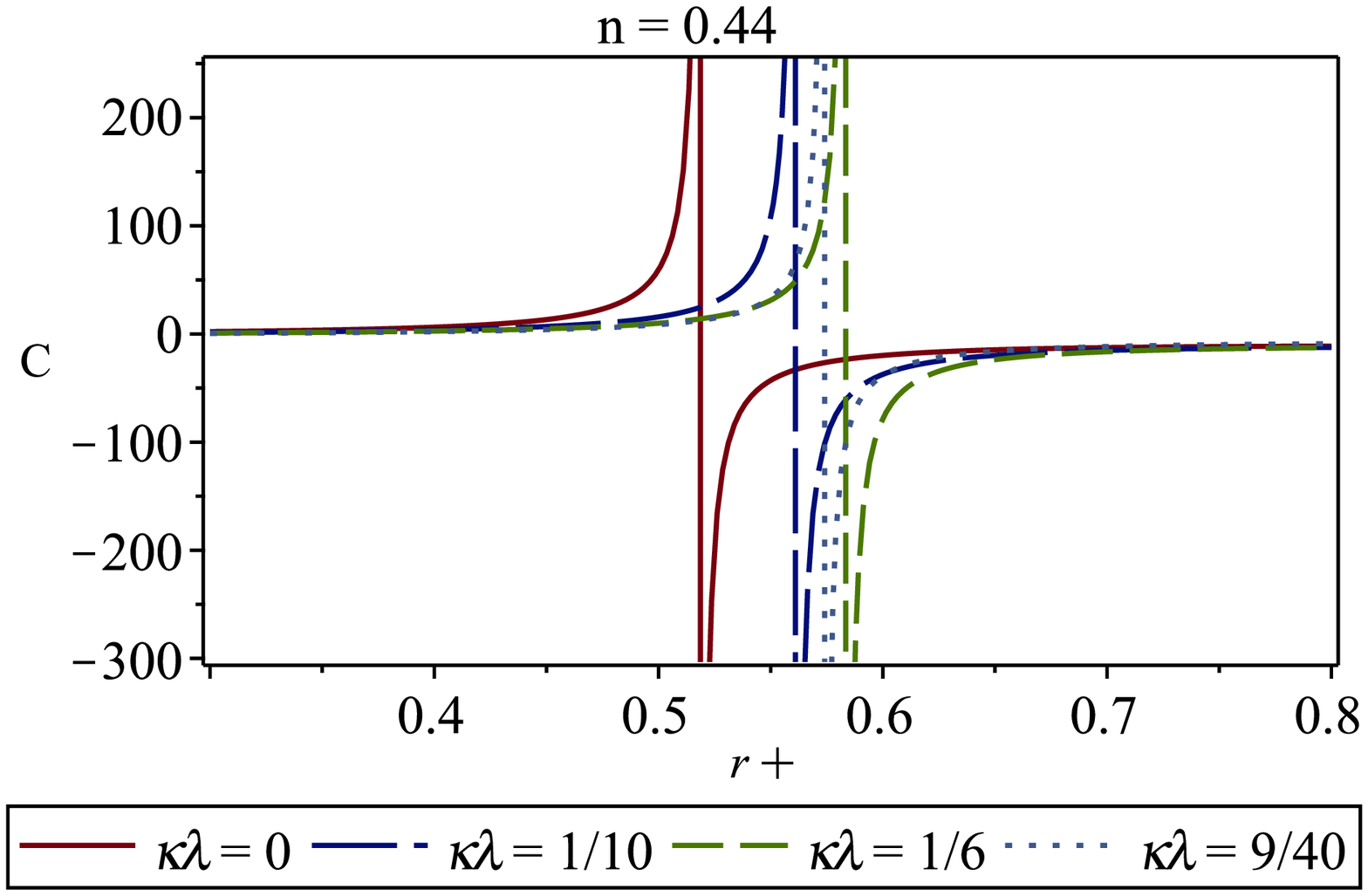}\\
		\includegraphics[scale=0.3]{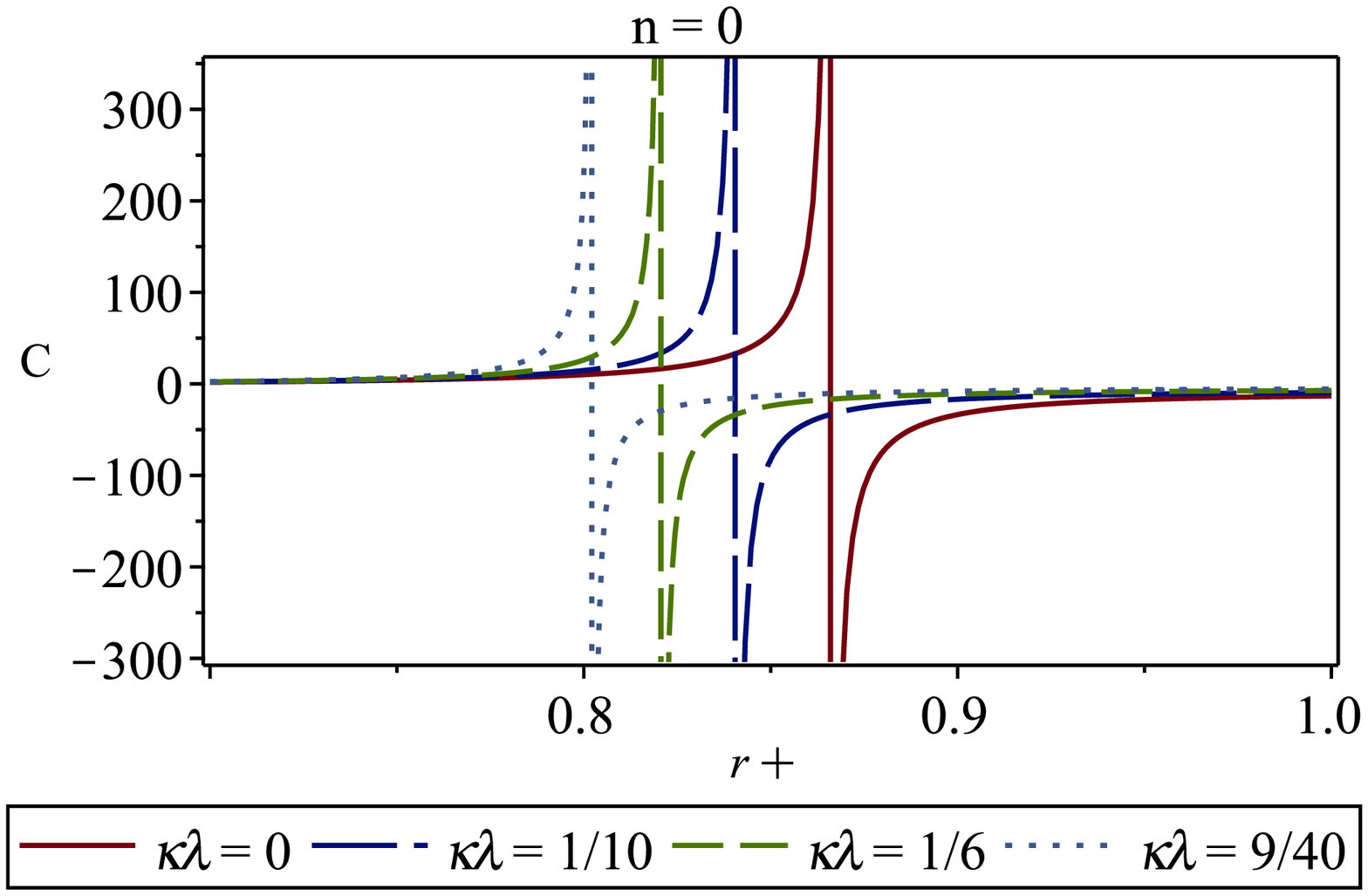}\includegraphics[scale=0.3]{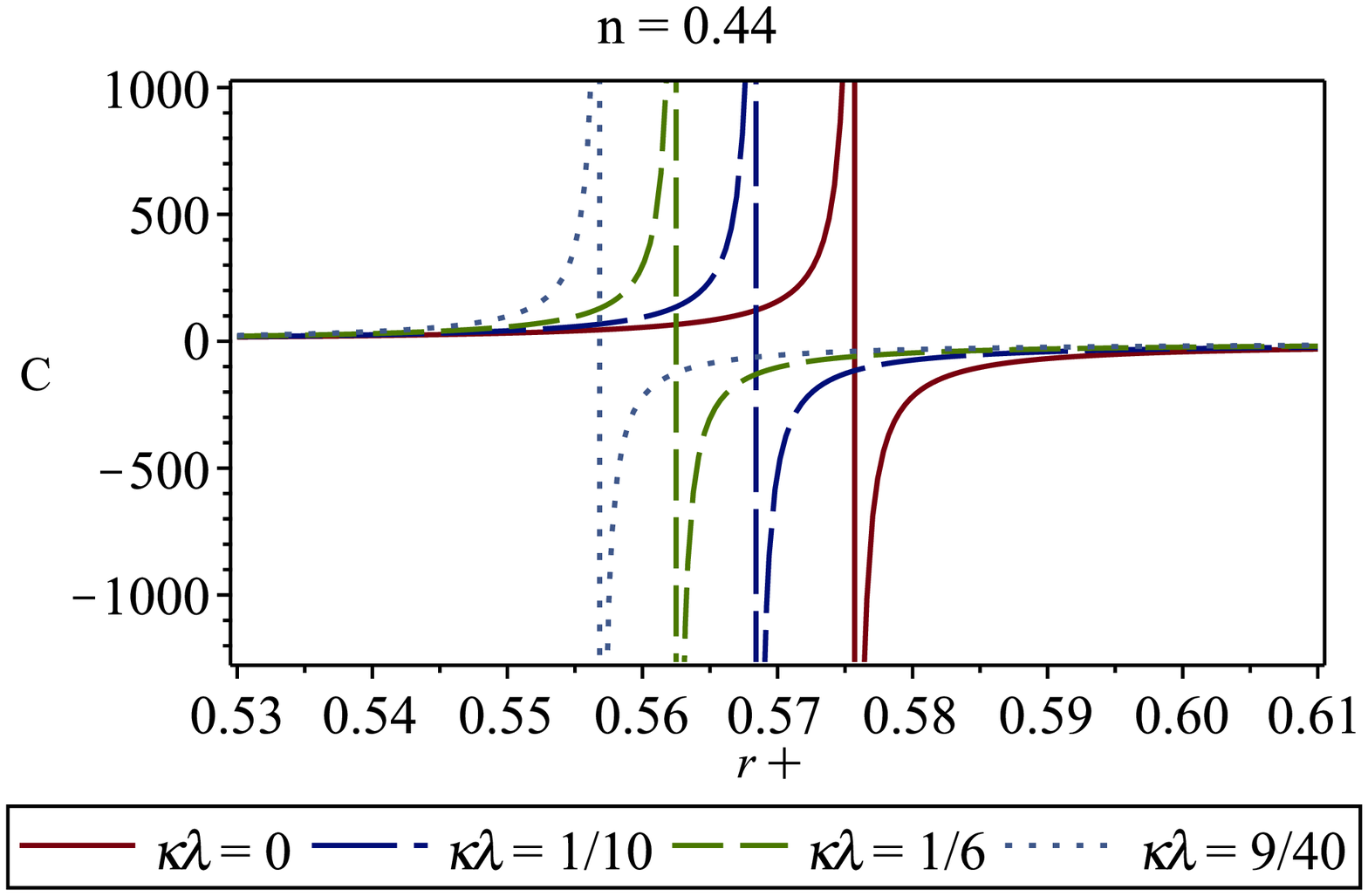}
		\caption{Plot showing the heat capacity of a KNN-R black hole horizon surrounded by dust field (top) and quintessence field (bottom) when the outer horizon $r_+$ varies. Here we use $Q=0.5$, $a\approx0$, and $N_s=0.1$.}\label{gambar:7}
	\end{center}
\end{figure}

Aside from the entropy of a black hole, Bekenstein-Hawking formula also gives the temperature of a horizon, related to its surface gravity, given by (in Planck units) \cite{25,53}
\begin{equation}
T=\frac{\kappa}{2\pi}=\frac{\Delta'(r_+)}{4\pi(r_+^2+a^2+n^2)}\approx\frac{\Delta'(r_+)}{4\pi(r_+^2+n^2)}.
\end{equation}
For a slowly-rotating KNN-R black hole, the temperature will have the form
\begin{equation}
T\approx\frac{1}{4\pi(r_+^2+n^2)}\bigg[2(r_+-M)+\frac{N_s(\zeta-2)r_+}{(r_+^2+n^2)^\frac{\zeta}{2}}\bigg],
\end{equation}
where $M$ is the mass of a slowly-rotating KNN-R black hole,
\begin{equation}
M\approx\frac{1}{2r_+}[r_+^2-n^2+Q^2-N_s(r_+^2+n^2)^{\frac{2-\zeta}{2}}].\label{eq:56'}
\end{equation}
For the case of vanishing surrounding matter, \textit{i.e.} $N_s\rightarrow0$, we recover the Bekenstein-Hawking temperature for a Kerr-Newman-NUT black hole. We also get zero temperature for the extremal case. We can see how the temperature behaves when $r_+$ varies with different NUT $n$ and Rastall parameter in Fig. \ref{gambar:6}. The presence of NUT $n$ parameter makes the temperature slightly increases. As the black hole evaporates, the horizon radius started to decrease and the temperature rises until it reaches a certain maximum point and continuously falls to absolute zero when it reaches an extremity, \textit{i.e.} when there is only one horizon. When KNN-R black hole is surrounded by dust field, the temperature reaches a certain asymptotic value if $r_+$ is quite large (except when $\kappa\lambda=9/40$). On the other hand, when the black hole is surrounded by quintessence field, the temperature falls again to absolute zero after the outer horizon reaches a certain point.\\
\indent Heat capacity can be obtained using the formula given by \cite{1,25}
\begin{equation}
C=\frac{dM}{dT}\bigg|_{r=r_+}=\frac{dM/dr_+}{dT/dr_+}.\label{eq:55}
\end{equation}
Hence, from Eq. (\ref{eq:55}) and (\ref{eq:56'}), we have the explicit form of the heat capacity,
\begin{equation}
C\approx\frac{2\pi(r_+^2+n^2)^2(r_+^2+N_s\alpha(r_+))}{r_+^2+(Q^2-n^2)(1+r_+^2)+N_s\beta(r_+)},
\end{equation}
where $\alpha(r_+)$ and $\beta(r_+)$ are functions defined by
\begin{equation}
\alpha(r_+)=(r_+^2(\zeta-1)+n^2)(r_+^2+n^2)^\frac{-\zeta}{2}+n^2-Q^2,
\end{equation}
\begin{equation}
\beta(r_+)=[(r_+^4\zeta(\zeta-2)-(r_+^2+n^2)^2/2)(r_+^2+n^2)^{-1}+(r_+^4(1-\zeta)-n^2)](r_+^2+n^2)^{\frac{-\zeta}{2}}.
\end{equation}
The behavior of heat capacity can be useful to investigate the thermodynamic stability of a black hole. A black hole is said to be thermodynamically stable when the heat capacity is positive-valued, \textit{i.e.} when $C>0$ and is thermodynamically unstable when $C<0$. Fig. \ref{gambar:7} shows regions where the black hole is thermodynamically stable. The value varies for different values of Rastall parameter, and the existence of NUT parameter also affects the stable region.
\section{Equatorial Circular Orbits Analysis}
\begin{figure}
	\centering
	\includegraphics[scale=0.3]{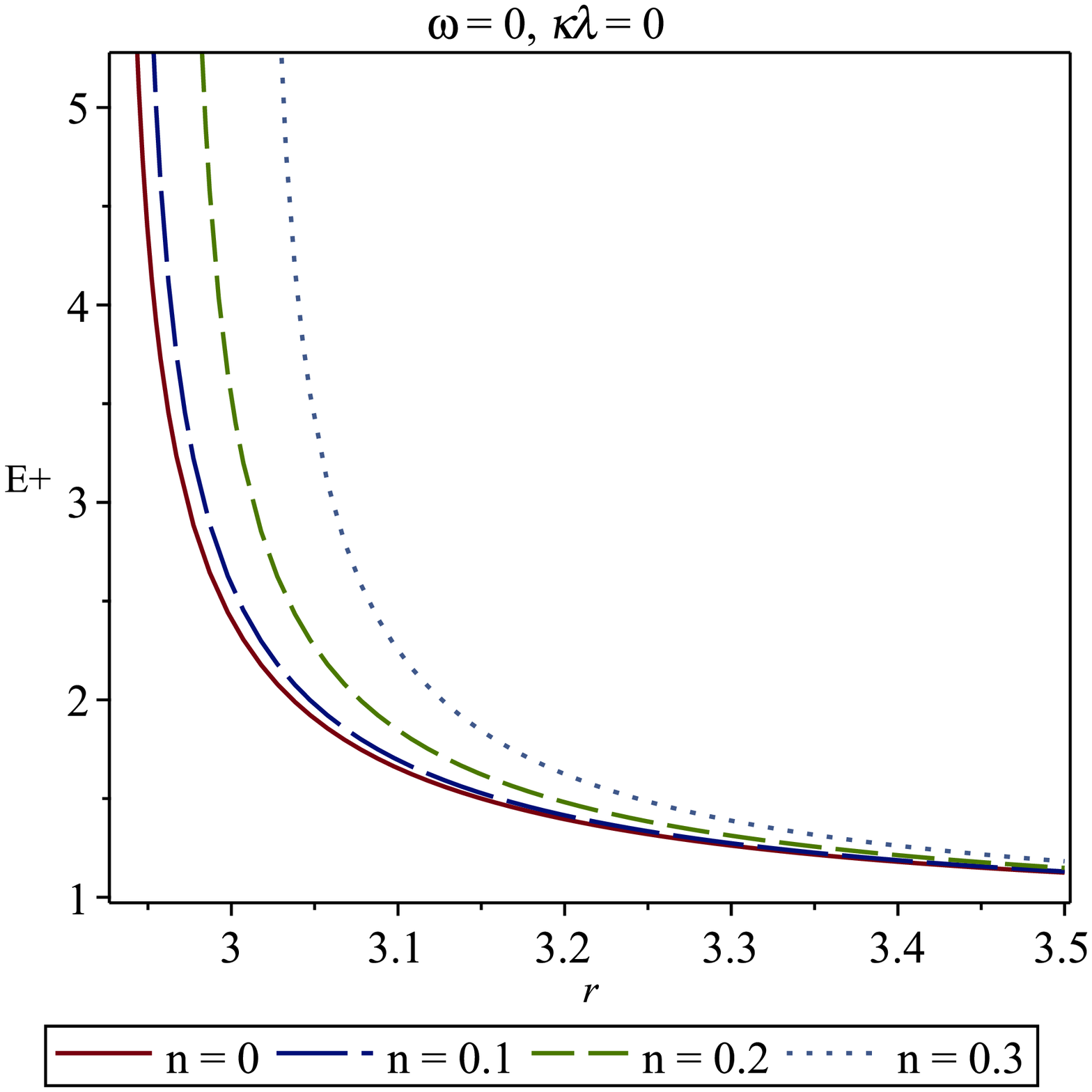}
	\includegraphics[scale=0.3]{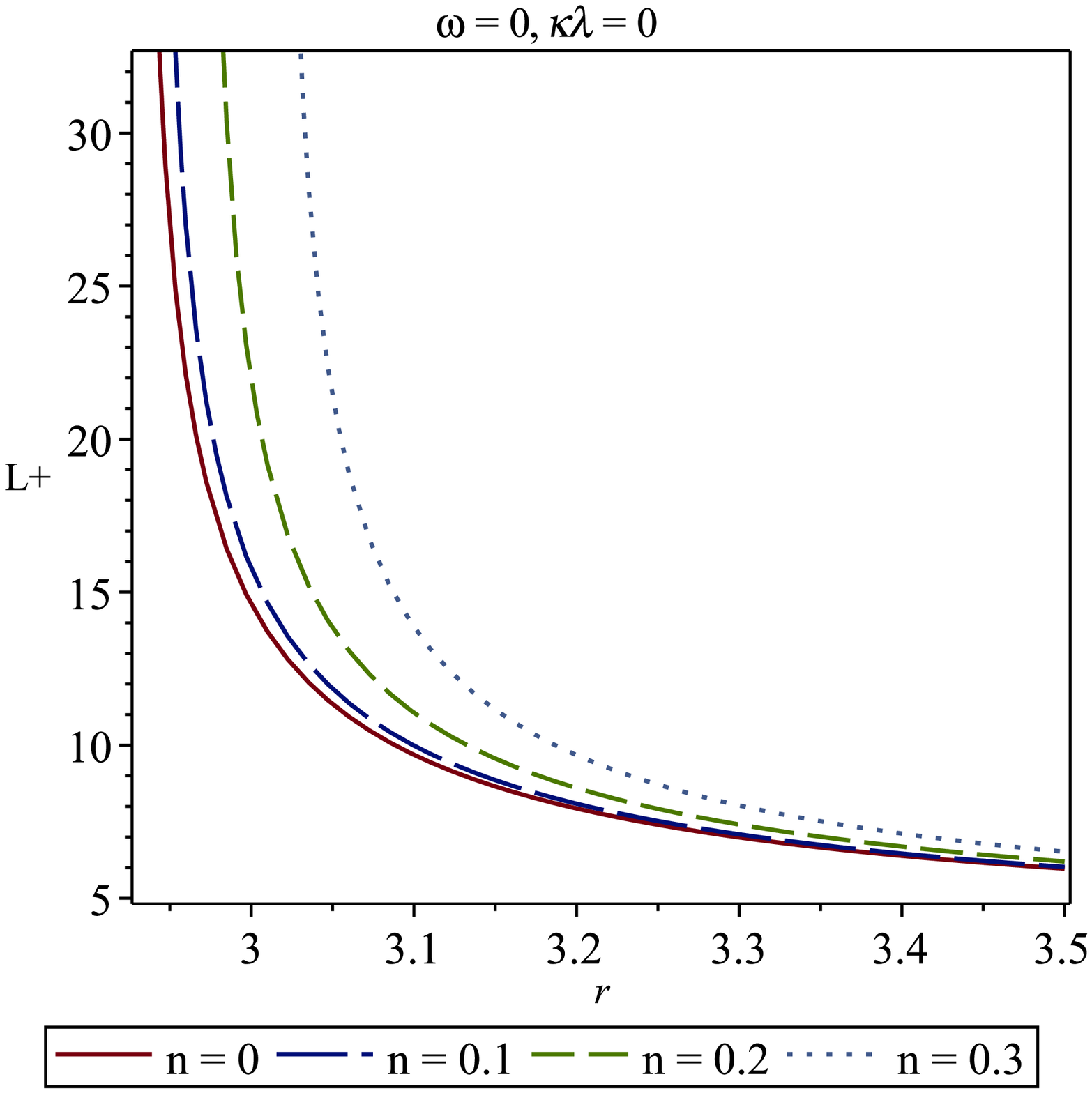}
	\caption{\label{gambar:4} Plot of the co-rotating energy $E_+$ and angular momentum $L_{z+}$ with respect to r for a KNN-R black hole surrounded by dust field with $M=1, a=0.9, Q=0.5, \kappa\lambda=0$, and $N_s=-0.01$. For the same amount of energy and angular momentum, a time-like particle could revolve closer to the black hole when the NUT parameter decreases.}
\end{figure}

\begin{figure}
	\centering
	\includegraphics[scale=0.3]{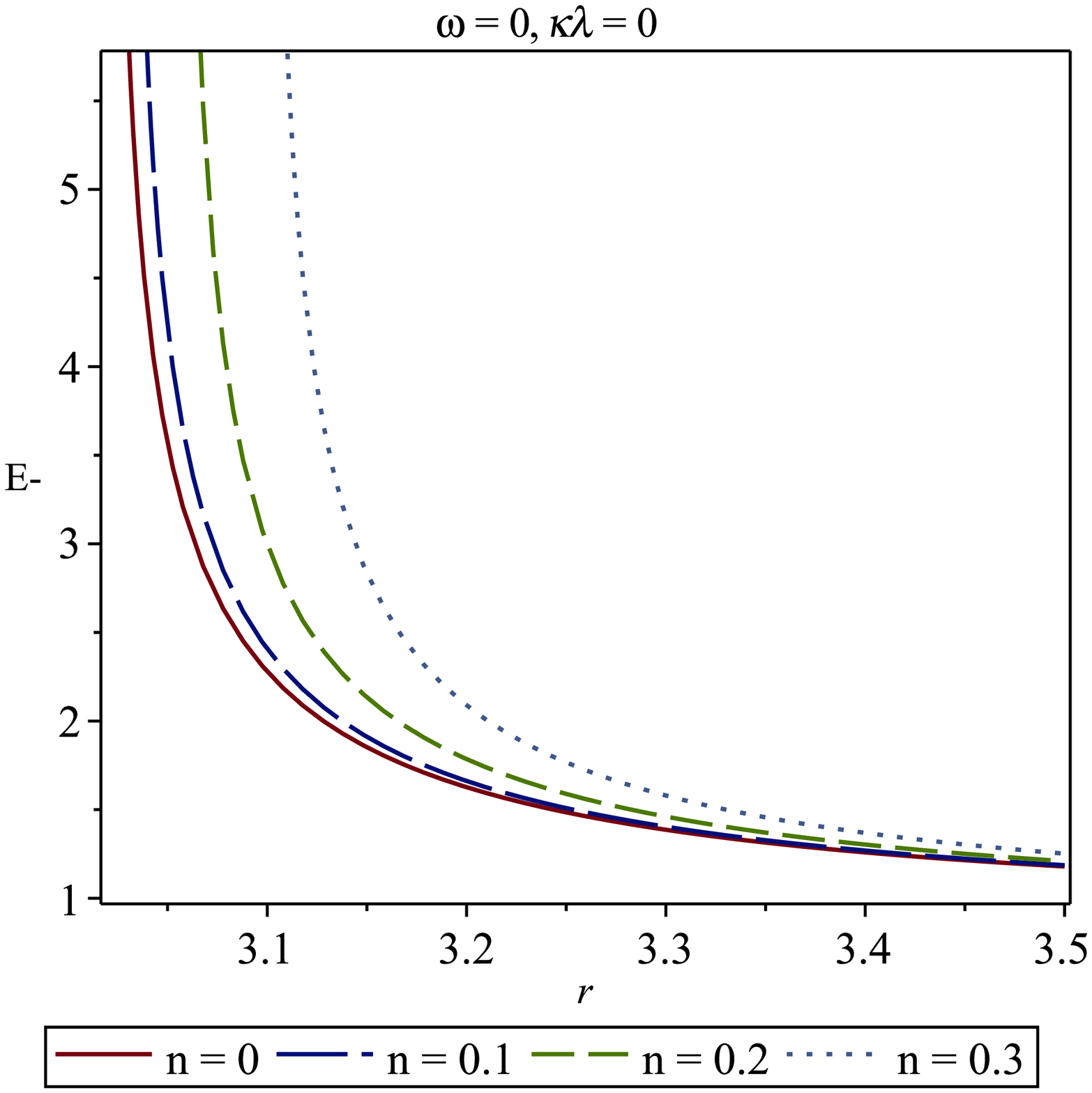}
	\includegraphics[scale=0.3]{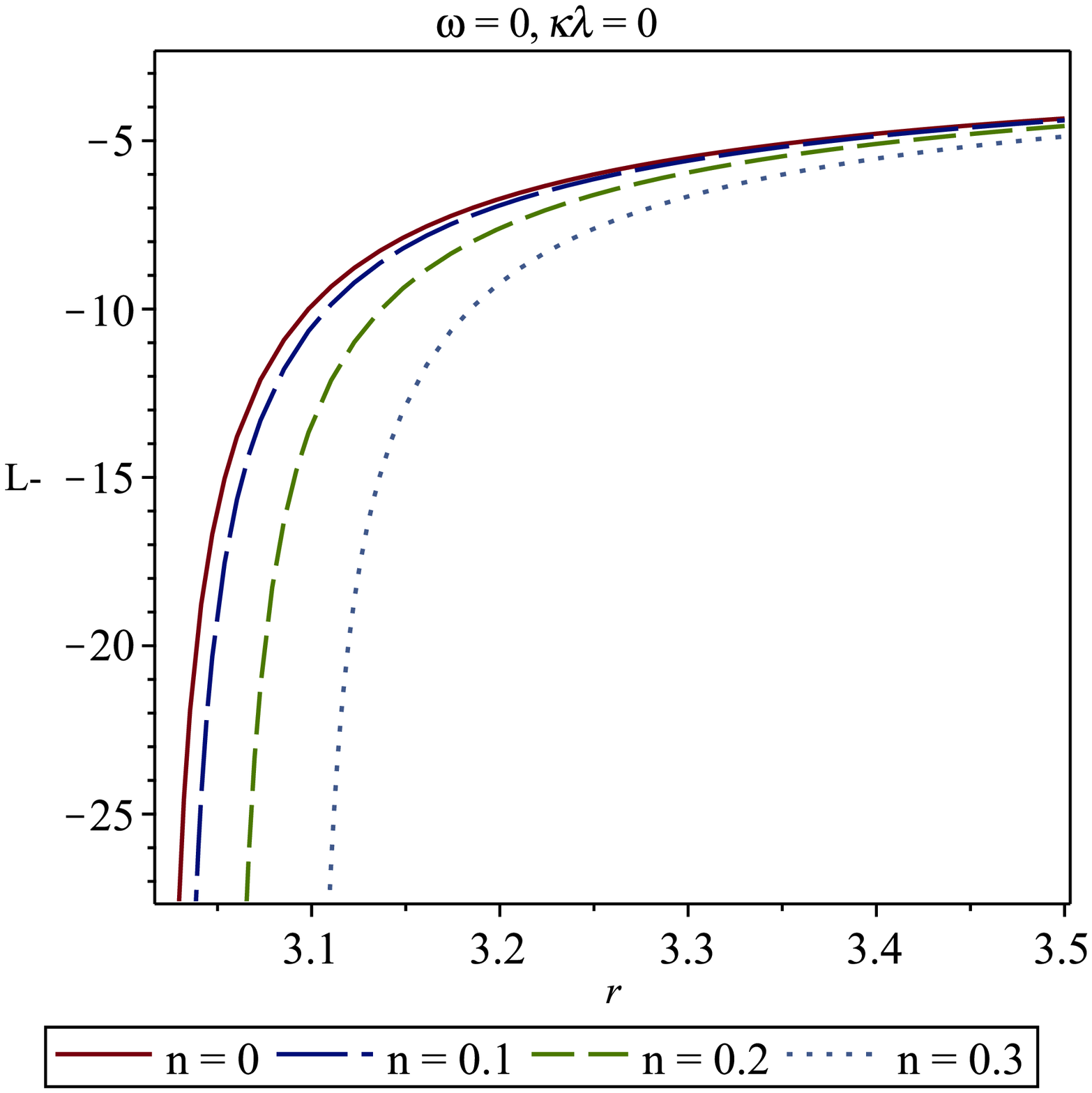}
	\caption{\label{gambar:5} Plot of the counter-rotating energy $E_-$ and angular momentum $L_{z-}$ with respect to r for a KNN-R black hole surrounded by dust field with $M=1, a=0.9, Q=0.5, \kappa\lambda=0$, and $N_s=-0.01$. This gives a similar result with the previous co-rotating plot.}
\end{figure}

In this section, we would like to investigate the motion of a particle (both massive and massless) orbiting the KNN-R black hole with quintessential background. In order to do so, first, we need to consider the Lagrangian for following particle that given by
\begin{equation}
\mathcal{L}=\frac{1}{2}g_{\mu\nu}\dot{x}^{\mu}\dot{x}^{\nu}.\label{eq:45}
\end{equation}
Here we use the dot notation as the derivative with respect to an Affine parameter $\tau$, $\dot{x}^{\mu}\equiv dx^{\mu}/d\tau$. Metric tensor $g_{\mu\nu}$ used in Eq. (\ref{eq:45}) is given by metric elements in Eq. (\ref{eq:40}), \textit{i.e.} line element for the KNN-R black hole. Since we are only interested in circular equatorial geodesic, we will use $\theta=\pi/2$ and $\dot{r}=\dot{\theta}=0$ afterward.\\
\indent We have mentioned before that KNN-R black hole, as an axisymmetric black hole, will have symmetries under time and axial translation. Hence, we will have two conserved quantities, energy $E$ and angular momentum $L_z$ that can be derived straight from the Lagrangian as
\begin{align}
-E&=\frac{\partial\mathcal{L}}{\partial\dot{t}}=g_{tt}\dot{t}+g_{t\phi}\dot{\phi},\;\;\;\;\;
L_z=\frac{\partial\mathcal{L}}{\partial\dot{\phi}}=g_{t\phi}\dot{t}+g_{\phi\phi}\dot{\phi},\label{eq:56}
\end{align}
and the other momentum components are $p_r=g_{rr}\dot{r}$, $p_{\theta}=0$. The Hamiltonian
\begin{equation}
2\mathcal{H}=-E\dot{t}+g_{rr}\dot{r}^2+L_z\dot{\phi}
\end{equation}
is equal to $-1,0,1$ for a particle with a time-like, null-like, and space-like geodesics, respectively \cite{52,54}. Following Eq. (\ref{eq:56}), and assume that the particle follows a time-like geodesics, we will have
\begin{equation}
g_{rr}\dot{r}=\frac{g_{\phi\phi}E^2+2g_{t\phi}EL_z+g_{tt}L_z^2}{g_{t\phi}^2-g_{tt}g_{\phi\phi}}-1,
\end{equation}
and from the equation above, we can define the effective potential as
\begin{equation}
V_{eff}=\frac{g_{\phi\phi}E^2+2g_{t\phi}EL_z+g_{tt}L_z^2}{g_{t\phi}^2-g_{tt}g_{\phi\phi}}-1,
\end{equation}
and therefore having an explicit expression given by
\begin{equation}
V_{eff}(r)=\frac{((r^2+n^2+a^2)E-aL_z)^2-\Delta(r)((aE-L_z)^2+(r^2+n^2))}{\Delta(r)(r^2+n^2)}.\label{eq:52lagi}
\end{equation}
Here, $\Delta(r)=r^2+a^2-n^2-2Mr+Q^2-N_s(r^2+n^2)^{\frac{2-\zeta}{2}}$ is a function of $r$ only. Equatorial circular orbits occurs at the zeros and turning points of the effective potential, hence we have $\dot{r}=\dot{\theta}=0$ which indicates $V_{eff}=0$, and $\ddot{r}=\ddot{\theta}=0$ which requires $\partial_rV_{eff}=\partial_{\theta}V_{eff}=0$ \cite{52,54}. Equation (\ref{eq:52lagi}) indicates that the effective potential blows up ($V_{eff}\rightarrow\infty$) when $r$ approaches the horizon ($\Delta\rightarrow0$). Following Ref. \citen{52,54}, using geodesic equation and equatorial circular orbit condition, we also have an expression for $E$ and $L_z$, and is given by
\begin{align}
E_{\pm}&=-\frac{g_{tt}+g_{t\phi}\Omega^{circ.}_{\pm}}{\sqrt{-g_{tt}-2g_{t\phi}\Omega^{circ.}_{\pm}-g_{\phi\phi}(\Omega^{circ.}_{\pm})^2}},\label{eq:49}\\
L_{z\pm}&=\frac{g_{t\phi}+g_{\phi\phi}\Omega^{circ.}_{\pm}}{\sqrt{-g_{tt}-2g_{t\phi}\Omega^{circ.}_{\pm}-g_{\phi\phi}(\Omega^{circ.}_{\pm})^2}},\label{eq:50}
\end{align}
where $\Omega^{circ.}_{\pm}=\dot{\phi}/\dot{t}$ is the particle's angular velocity restricted by circular equatorial condition, and $\pm$ denotes that the particle could have a co-rotating ($\Omega^{circ.}_+$) and counter-rotating ($\Omega^{circ.}_-$) angular velocity with respect to ZAMO. An expression for the angular velocity $\Omega^{circ.}_{\pm}$ is given by \cite{52}
\begin{equation}
\Omega^{circ.}_{\pm}=\frac{-\partial_rg_{t\phi}\pm\sqrt{(\partial_rg_{t\phi})^2-\partial_rg_{\phi\phi}\partial_rg_{tt}}}{\partial_rg_{\phi\phi}}.
\end{equation}
A straightforward computation using KNN-R metric elements leads us to an explicit form for $\Omega^{circ.}_{\pm}$, given by
\begin{align}
\Omega^{circ.}_{\pm}&=\frac{a[2r(\Delta(r)-a^2)-\Delta'(r)(r^2+n^2)]}{a^2[2r(\Delta(r)-a^2)-\Delta'(r)(r^2+n^2)]+2r(r^2+n^2)^2}\nonumber\\
&\;\;\;\;\;\pm\frac{(r^2+n^2)\sqrt{2r\Delta'(r)(r^2+n^2)-4r^2(\Delta(r)-a^2)}}{a^2[2r(\Delta(r)-a^2)-\Delta'(r)(r^2+n^2)]+2r(r^2+n^2)^2},\label{eq:52}
\end{align}
From here, it is possible to compute an explicit formula for $E$ and $L_z$ by plugging $\Omega^{circ.}_{\pm}$ in Eq. (\ref{eq:52}) to Eq. (\ref{eq:49}) and (\ref{eq:50}), hence the results are quite tedious. We are going to show how $E$ and $L_z$ varies with respect to $r$ by making a plot, given by Fig. \ref{gambar:4}. It is shown that the NUT parameter gives us an important information. As the NUT parameter $n$ increases, a time-like co-rotating particle will have a greater equatorial circular orbit radius for a given energy. That contradicts the rotation parameter $a$ as shown in Ref. \citen{52}, because as the NUT parameter $n$ increases, we also have greater horizon radius, as given by Fig. \ref{gambar:3}.  Surprisingly, a time-like counter-rotating particle gives the exact same behavior from its counterparts, for a varying NUT parameter, as shown in Fig. \ref{gambar:5}. It is also shown that the energy of a co- and counter-rotating time-like particle approaches infinity at a certain point. This gives us information about the existing of a null circular orbit, as we will discuss further.
\subsection{Static Radius Limit for a Time-like Particle}
We already showed that energy $E$ and angular momentum $L_z$ of a massive particle on a circular equatorial orbit depends on $\Omega^{circ.}_{\pm}$. Hence, both energy $E$ and angular momentum $L_z$ will be real valued if
\begin{equation}
\Delta'(r)(r^2+n^2)-2r(\Delta(r)-a^2)\geq0\label{eq:53}
\end{equation}
is satisfied. When there is no NUT parameter, \textit{i.e.} $n\rightarrow0$, we recover the static radius limit condition found in Ref. \citen{54} for a rotating black hole. Equation (\ref{eq:53}) gives an existence of a static radius defined by $r_s$, where $r\leq r_s$ are always satisfied for a test time-like particle. At $r=r_s$, an object is located at an unstable-equilibrium position with $L_z=0$ \cite{52,54}.\\
\indent In the case of the KNN-R black hole metric given by (\ref{eq:40}), we can have an explicit formula for Eq. (\ref{eq:53}), given by
\begin{equation}
2r(Mr+2n^2-Q^2)-2Mn^2+N_s\zeta r(r^2+n^2)^{\frac{2-\zeta}{2}}\geq0.\label{eq:58}
\end{equation}
Interestingly, Eq. (\ref{eq:58}) does not depends on the rotation parameter $a$, as in the case of the rotating black hole with the quintessential energy \cite{54}. We would like to see the values of $r_s$ for different values of $n$ and $\kappa\lambda$, as shown in Table \ref{tabel:3}. 
\begin{table}[H]
		\tbl{Table showing the static radius $r_s$ for different values of $n$ and $\kappa\lambda$. Here we take $M=1, Q=1, \omega=-2/3$ (quintessence field), $N_s=0.01$.}
	{\begin{tabular}{@{}cccccc@{}}
			\toprule
			n & $\kappa\lambda=0$ & $\kappa\lambda=1/10$ & $\kappa\lambda=1/6$ & $\kappa\lambda=1/4$ \\\colrule
			&  $r_s$ & $r_s$ & $r_s$ & $r_s$ \\ \colrule
			0.1 & 13.62302 & 8.10086 & 5.97651 & 4.25072  \\ 
			0.2 & 13.65359 & 8.12708 & 5.99994 & 4.27077  \\ 
			0.3 & 13.70408 & 8.17009 & 6.03816 & 4.30314  \\ 
			0.4 & 13.77379 & 8.17009 & 6.09003 & 4.34644  \\ 
			0.5 & 13.86184 & 8.30246 & 6.15413 & 4.39904  \\
			0.6 & 13.86184 & 8.38922 & 6.22893 & 4.45923  \\ 
			0.7 & 14.08853 & 8.48780 & 6.31286 & 4.52541  \\
			0.8 & 14.22474 & 8.59673 & 6.40442 & 4.59612  \\
			0.9 & 14.37449 & 8.71462 & 6.50224 & 4.67012  \\
			1 & 14.53651 & 8.84017 & 6.60510 & 4.74637  \\ \botrule
		\end{tabular}\label{tabel:3}}
\end{table}
When the KNN-R black hole is surrounded by the quintessence field, an increasing $n$ gives an increasing $r_s$ as well, for various value of $\kappa\lambda$, \textit{e.g.} we have $r_s=13.62302$ for $n=0.1$ while $r_s=13.65359$ for $n=0.2$, with $\kappa\lambda=0$. An increasing Rastall parameter makes the static radius $r_s$ decreases, \textit{e.g.} when $\kappa\lambda=0$ and $\kappa\lambda=1/10$, we have $r_s=14.53651$ and $r_s=8.84017$, respectively, for $n=1$.
\subsection{Null Equatorial Circular Orbit}
\indent For null particles, such as photon, the corresponding velocity vector $u^{\mu}=dx^{\mu}/d\tau$ satisfies $u^{\mu}u_{\mu}=0$. This also indicates that the Lagrangian, given by Eq. (\ref{eq:45}), is equal to zero. Therefore, from the equatorial circular orbit condition, we have an equation that gives the equatorial null circular orbit, as\cite{52}
\begin{equation}
g_{tt}+2g_{t\phi}\Omega^{circ.}_{\pm}+g_{\phi\phi}(\Omega^{circ.}_{\pm})^2=0.\label{eq:59}
\end{equation}
This condition happens when the energy $E$ of a time-like particle approaches infinity. Hence, for most cases, $r>r_{\gamma}$ is always satisfied for a time-like particle, where $r_{\gamma}$ is defined as the equatorial null circular orbit, \textit{i.e.} the root(s) of equation (\ref{eq:59}). Here we are only considering the co-rotating null equatorial circular orbit, as the zeros of
\begin{align}
\label{eq:73}0&=2a^2[\Delta(r)-(r^2+n^2+a^2)][2r(\Delta(r)-a^2)-\Delta'(r)(r^2+n^2)](a^2[2r(\Delta(r)\\\nonumber
&\;\;\;-a^2)-\Delta'(r)(r^2+n^2)]+2r(r^2+n^2)^2)+((r^2+a^2+n^2)^2-a^2\Delta(r))(a^2\\\nonumber
&\;\;\;\times[2r(\Delta(r)-a^2)-\Delta'(r)(r^2+n^2)]^2+(r^2+n^2)^2[2r\Delta'(r)(r^2+n^2)-4r^2\\\nonumber
&\;\;\;\times(\Delta(r)-a^2)])+\{a(\Delta(r)-(r^2+a^2+n^2))(r^2+n^2)[a^2[2r(\Delta(r)-a^2)\\\nonumber
&\;\;\;-\Delta'(r)(r^2+n^2)]+2r(r^2+n^2)^2]+2a(r^2+n^2)((r^2+a^2+n^2)-a^2\Delta(r))\\\nonumber
&\;\;\;\times[2r(\Delta(r)-a^2)-\Delta'(r)(r^2+n^2)]\}\sqrt{2r\Delta'(r)(r^2+n^2)-4r^2(\Delta(r)-a^2)},
\end{align}
where the dependence of other parameters such as $Q, N_s, \omega, \kappa\lambda,$ and $M$ contained implicitly on the $\Delta(r)$ function. From Eq. (\ref{eq:73}), we can see that the condition (\ref{eq:53}) also needs to be satisfied for the null equatorial circular orbit to be real-valued, \textit{i.e.} $r_{\gamma}<r_s$, thus again confirms that $r>r_{\gamma}$. The results for various NUT and rotation parameter are given by Table \ref{tabel:5} for a KNN-R black hole surrounded by dust field. As we can see, the NUT parameter makes $r_{\gamma}$ increases, while the rotation parameter makes $r_{\gamma}$ decreases for the few first, then starting to increase at some point. This also corresponds with the increasing and decreasing horizon with respect to those parameters. For example, when $a=0$ we have $r_{\gamma}=2.853297101$ and $r_{\gamma}=2.901130409$ for $n=0$ and $n=0.22$, respectively.
\begin{table}[H]
	\tbl{\label{tabel:5}The values of $r_{\gamma}$ for $\omega=0$ (dust field), $\kappa\lambda=1/6$ with variations of NUT and rotation parameters. Here we use $M=1, Q=0.5, N_s=0.01$.}
	{\begin{tabular}{@{}cccccc@{}}
			\hline
			& $n=0$ & $n=0.22$  & $n=0.44$  & $n=0.66$ & $n=0.88$\\ 
			\hline
			a & $r_{\gamma}$ & $r_{\gamma}$ & $r_{\gamma}$ & $r_{\gamma}$ & $r_{\gamma}$\\ 
			\hline
			0 &  2.853297101  &  2.901130409 &  3.036616821 & 3.241211157 & 3.495387746 \\ \hline
			0.1 &  2.824983652  &  2.874246358 &  3.013240269 & 3.221963944 & 3.479920516 \\ \hline 
			0.2 &  2.801408682  &  2.852074065 &  2.994452148 & 3.207054944 & 3.468459685 \\ \hline 
			0.3 &  2.785137766  &  2.837023148 &  2.982290865 & 3.198108366 & 3.462273335 \\ \hline 
			0.4 &  2.778990935  &  2.831717688 &  2.978930901 & 3.196821800 & 3.462666978 \\ \hline 
			0.5 &  2.785769670  &  2.838754301 &  2.986508067 & 3.204854444 & 3.470915179 \\ 
			\hline
	\end{tabular}}
\end{table}
\subsection{Innermost Stable Circular Orbit (ISCO)}
\begin{figure}
	\begin{center}
		\includegraphics[scale=0.3]{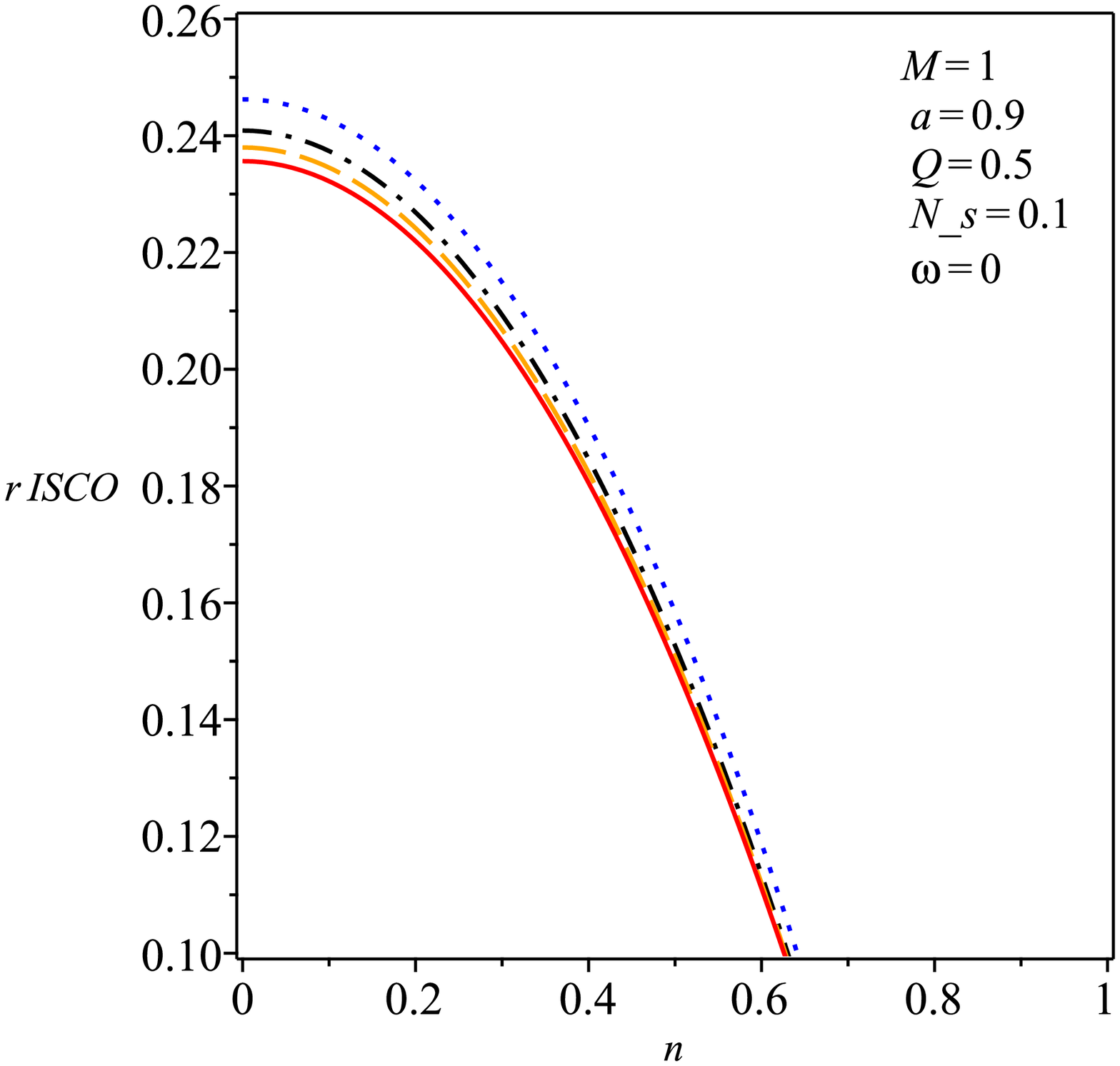}\includegraphics[scale=0.3]{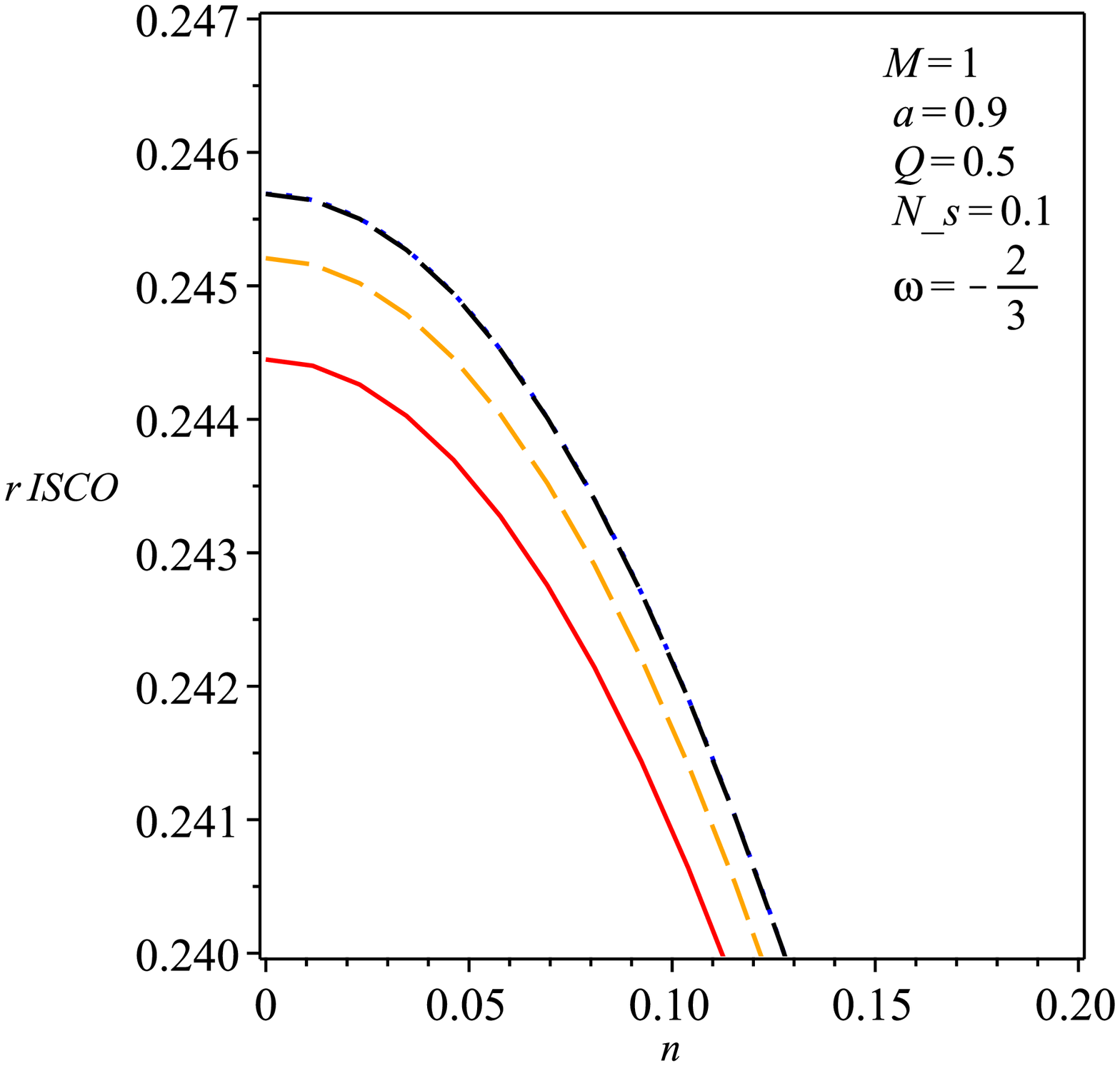}
		\caption{Plot showing the behavior of $r_{ISCO}$ when the NUT parameter $n$ varies for different value of Rastall parameter for a KNN-R black hole surrounded by dust (left) and quintessence (right) field. Here \textit{blue-dot}, \textit{black-dot-dashed}, \textit{orange-dashed}, and \textit{red-solid} line represents $\kappa\lambda=1/4,1/6,1/10,0$, respectively.}\label{gambar:8}
	\end{center}
\end{figure}

Another important aspect of a particle geodesic around black hole is the Innermost Stable Circular Orbit (ISCO). When
\begin{equation}
\frac{\partial^2V_{eff}}{\partial r^2}=0,
\end{equation}
the equatorial circular orbit becomes $r_{ISCO}$ and every orbit that smaller than $r_{ISCO}$ will be unstable \cite{52}. The inner radius of an accretion disk around a black hole is assumed as the $r_{ISCO}$ according to the Novikov-Thorne model \cite{52}. From Eq. (\ref{eq:52lagi}) and using $V_{eff}=0$, $r_{ISCO}$ is then the zeros of
\begin{equation}
E^2(12r^2+a^2+n^2)-aL_zE-\Delta''(r)[(aE-L_z)^2+(r^2+n^2)]-4\Delta'(r)r-2\Delta(r)=0,
\end{equation}
where the double prime notation denotes the second derivative with respect to $r$. As we can see from Fig. \ref{gambar:8}, $r_{ISCO}$ is increased when the NUT parameter decreases, for both dust and quintessence field case.
\section{Conclusions and Discussions}
We obtained the Kerr-Newman-NUT black hole solution in Rastall theory of gravity (KNN-R black hole) using Newman-Janis algorithm as the generalized version of the solution found in Ref. \citen{1}. In this work, we have found that the horizon of the KNN-R black hole is $\theta$ dependence, \textit{i.e.} the shape of the horizon is not perfectly spherical. We studied the behavior of the horizon and ergosurface of KNN-R black hole in section 3. The horizon can have more than one solution, \textit{e.g.} inner horizon $r_-$, outer horizon $r_+$, cosmological horizon $r_q$, or even more horizons. Analytical solution of the horizon is shown in Table \ref{tabel:1} and the behavior of the $\Delta$ function is shown in Fig. \ref{gambar:1} and Fig. \ref{gambar:2}. The new NUT parameter affects the shape and of the horizon and ergosurface, as shown in Table \ref{tabel:2} and Fig. \ref{gambar:3}, that when the NUT parameter increases, the radius of the horizon also became slightly bigger.\\
\indent In section 4, we studied the thermodynamic properties of the KNN-R black hole. We investigate the black hole entropy using the Bekenstein-Hawking formula. Here we assume that the black hole is slowly-rotating to get rid with the $\theta$ dependence. Area and entropy of KNN-R black hole have been found in Eq. (\ref{eq:49'}) and (\ref{eq:entropi}), and coincides with the entropy of a slowly-rotating Kerr-Newman-NUT black hole \cite{46} with the Rastall parameter affect the horizon radius $r_+$ implicitly. We also see how the black hole temperature and heat capacity varies with respect to the outer horizon in Fig. \ref{gambar:6} and \ref{gambar:7}. The thermodynamic stability of the black hole also has been studied in this work from the behavior of the heat capacity. Existence of the NUT parameter affects the thermodynamically stable region.\\
\indent In this work, aside from Ref. \citen{1}, we also studied some geodesic properties of this black hole, \textit{i.e.} the equatorial circular orbit. From the Hamiltonian formulation, we derived the effective potential, as shown in Eq. (\ref{eq:52lagi}), and the angular velocity of a time-like particle with respect to ZAMO, as shown in Eq. (\ref{eq:52}). From that, we obtained the energy and angular momentum of a time-like particle and it turns out that when the NUT parameter increased, the particle's equatorial circular orbit is also increased. We investigated the static radius limit in Table \ref{tabel:3}, which showed that the increasing NUT parameter causes an increasing static radius limit. The null equatorial circular orbits are also obtained in Table \ref{tabel:5} for various values of rotation and NUT parameter. Finally, we obtained the innermost stable circular orbit (ISCO) and plotted the results in Fig. \ref{gambar:8}. When the NUT parameter decreases, we found that $r_{ISCO}$ increases.
\section*{Acknowledgments}

F.P.Z. and G.H. would like to thank Kemenristek DIKTI Indonesia for financial supports. H.L.P. and M.F.A.R.S. would like to thank the members of Theoretical Physics Groups of Institut Teknologi Bandung for the hospitality. 


\end{document}